\documentclass[twocolumn,aps,ams,showpacs,superscriptaddress]{revtex4}

\usepackage{graphicx,color,soul}
\usepackage{latexsym}
\usepackage{amsmath,amssymb}        
\usepackage[draft=false]{hyperref}
\usepackage{mathrsfs}

\usepackage{mathrsfs}

\begin{document}


\title{Rotation curves of rotating galactic BEC dark matter halos}


\author{F. S. Guzm\'an}
\affiliation{Department of Physics and Astronomy, 
	University of British Columbia.
	6224 Agricultural Road, Vancouver BC, Canada, V6T1Z1.}
\affiliation{Instituto de F\'{\i}sica y Matem\'{a}ticas, Universidad
              Michoacana de San Nicol\'as de Hidalgo. Edificio C-3, Cd.
              Universitaria, 58040 Morelia, Michoac\'{a}n,
              M\'{e}xico.}

\author{F. D. Lora-Clavijo}
\affiliation{Instituto de Astronom\'ia, Universidad Nacional Aut\'onoma de M\'exico, 
	AP 70-264, Distrito Federal 04510, M\'exico.}

\author{J. J. Gonz\'alez-Avil\'es}
\affiliation{Instituto de F\'{\i}sica y Matem\'{a}ticas, Universidad
              Michoacana de San Nicol\'as de Hidalgo. Edificio C-3, Cd.
              Universitaria, 58040 Morelia, Michoac\'{a}n,
              M\'{e}xico.}
 
\author{F. J. Rivera-Paleo}
\affiliation{Instituto de F\'{\i}sica y Matem\'{a}ticas, Universidad
              Michoacana de San Nicol\'as de Hidalgo. Edificio C-3, Cd.
              Universitaria, 58040 Morelia, Michoac\'{a}n,
              M\'{e}xico.}


\date{\today}


\begin{abstract}
We present the dynamics of rotating Bose Condensate galactic dark matter halos, made of an ultralight spinless boson. We restrict to the case of adding axisymmetric rigid rotation to initially spherically symmetric structures and show there are three regimes: i) small angular momentum, that basically retains the drawbacks of spherically symmetric halos related to compactness and failure at explaining galactic RCs, ii) an intermediate range of values of angular momentum that allow the existence of long-lived structures with acceptable RC profiles, and iii) high angular momentum, in which the structure is dispersed away by rotation. We also present in detail the new code used to solve the Gross-Pitaevskii Poisson system of equations in three dimensions.
\end{abstract}




\maketitle


\section{Introduction}

The nature of dark matter is one of the most important quests in theoretical physics nowadays. Candidates are restricted by one or another observation. Nevertheless, the most successful cosmological model is the $\Lambda$CDM, which does not assume any particular dark matter particle.   An interesting dark matter candidate that escapes to stringent limitations is a spinless ultralight boson. The idea originates at cosmic scale, where a scalar field in effective theories is assumed to play the role of dark matter, and in fact what happens is that such candidate mimics CDM at cosmic scale, and most importantly, the mass power spectrum of structures fits better than CDM by fixing the mass of the scalar field to an ultralight value of $m\sim 10^{-22} - 10^{-23}eV$ \cite{MatosUrena2000}. Additionally, if such boson exists, the condensation temperature would be of the order of $T_c \sim $TeV for boson masses of $m\sim 10^{-22}$eV \cite{RoblesMatos2012}, which would allow the early formation of structures.

The idea of the scalar field in an affective theory works fine at cosmic scales, however once the structures are assumed to evolve under their own gravitational field, in a low energy and weak gravitational field limit, the Einstein-Klein-Gordon system is replaced by the time-dependent Schr\"odinger-Poisson (SP) system of equations \cite{GuentherThesis,Choi,Wai,BernalTesis}.

It happens that the SP system is interpreted as the equations ruling the evolution of a self-gravitating Bose Condensate, in the mean field approximation, where the trap of the condensate is the potential due to the gravitational field produced by the density of the condensate itself  \cite{GuzmanUrena2006,ChavanisTF} and Schr\"odinger equation is the Gross-Pitaevskii equation describing the Bose gas \cite{GrossPitaevskii}. The coupled system is then called the Gross-Pitaevskii-Poisson (GPP) system of equations. These equations are

\begin{eqnarray}
i\hbar \frac{\partial \tilde{\Psi}}{\partial \tilde{t}} &=& -\frac{\hbar^2}{2m}\tilde{\nabla}^2 \tilde{\Psi} + \tilde{V}\tilde{\Psi} +\frac{2\pi \hbar^2 \tilde{a}}{m^2} |\tilde{\Psi}|^2\tilde{\Psi}, \nonumber\\
\tilde{\nabla}^2 \tilde{V} &=& 4 \pi G m |\tilde{\Psi}|^2, \label{eq:SPcompleta}
\end{eqnarray}

\noindent where in general $\tilde{\Psi} = \tilde{\Psi}(\tilde{t},\tilde{{\bf x}})$, $m$ is the mass of the boson, $\tilde{V}$ is the gravitational potential acting as the condensate trap and $\tilde{a}$ is the scattering length of the bosons. 
In spherical symmetry, when the wave function is assumed to depend harmonically in time, the wave function can be written as $\tilde{\Psi} = e^{i\tilde{E}\tilde{t}/\hbar}\psi(\tilde{r})$, and the system of equations above reduces to a Sturm-Liouville eigenvalue problem for $\psi(\tilde{r})$, provided boundary conditions on the wave function. Solutions to such eigenvalue problem have been constructed numerically for instance in \cite{Ruffini,Gleiser1998,SinJin,GuzmanUrena2004,GuzmanUrena2006}. 
There are various aspects of the dynamical behavior of ground state solutions of this type, which  could be potentially related to astrophysical phenomena. For instance, the virialization time of structures is short after the turnaround  point \cite{GuzmanUrena2003}; BEC structures ruled by the GPP system of equations show a solitonic behavior, which could be consistent with the Bullet Cluster observation \cite{BernalGuzman2006b,Choi2008,GonzalezGuzman2011}; finally, one of the most appealing properties of these solutions is that ground state solutions are stable under very general perturbations \cite{BernalGuzman2006a}, and they are late-time attractors of general initial density fluctuations \cite{GuzmanUrena2006,BernalGuzman2006a}.

The size and mass scales of astrophysical systems associated to the equilibrium solutions of the GPP system depend on the mass of the boson considered. In the case of galaxy modeling, for an ultralight  boson $m \sim 10^{-23}$eV, equilibrium configurations are unable to explain the galactic rotation curves (RCs) because they are considerably compact, which in turn implies that RCs are Keplerian at short distances from the center of a galaxy halo. Two possibilities to rescue these equilibrium configurations as galactic halos are: i) that excited state solutions of the GPP system could be the halos, in fact thwy show desirable RCs \cite{SinJin,GuzmanUrena2004}, unfortunately these were shown to be unstable and therefore useless as galactic halos \cite{SinJin,GuzmanUrena2006} and ii) a new possibility proposes that galactic halos are a mixture of ground and excited states that unlike the pure excited states are stable and show promising RCs \cite{BernalUrena2012}; the question this mixed state configurations has to answer is how, being the critical temperature for condensation high of the order of TeV, at the current universe's temperature there are still a considerable amount of bosons in excited states.

An alternative view of BEC dark matter is found in the Thomas-Fermi (TF) limit of the GPP system, where the self-interaction between pairs of boson dominates over the kinetic terms in Schr\"odinger (Gross-Pitaevskii) equation. The model was presented in \cite{BohemerHarko2007} and the resulting density profile is of the type $\tilde{\rho}_{BEC}(r) = \tilde{\rho}_{BEC}^{c} \frac{\sin(\pi \tilde{r}/\tilde{R})}{\pi \tilde{r}/\tilde{R}}$, where $\tilde{R}$ is the radius at which $\tilde{\rho}_{BEC}(\tilde{R})=0$, defined as the size of the galaxy, and $\tilde{\rho}_{BEC}^{c}$ is the central density of dark matter; with these two parameters, a number of galactic RCs were fitted. Two drawbacks of this model are: 1) the characteristic radius is given by $\tilde{R}=\sqrt{\frac{\hbar^2 \tilde{a}}{Gm^3}}$, where $\tilde{a}$ is the self-interaction parameter between pairs of bosons and $m$ is the boson mass, and therefore, given a boson dark matter candidate with given  mass and self-interaction, the value of $\tilde{R}$ would be fixed for all galaxies and cannot be a free fitting parameter; a different way of looking at this problem is assuming $R$ can be different for different galaxies, then for a given boson mass $m$ every galaxy will require a different value of $a$, which in turn means that bosons interact differently in different galaxies;  2) the mass of the boson found in \cite{BohemerHarko2007} is of the order of meV-eV, which is inconsistent with the boson mass fitting the mass power spectrum, which requires the boson to be ultralight \cite{MatosUrena2000}.

An attempt to avoid the second inconsistency was presented in \cite{RoblesMatos2012}, where the same model under the TF approximation was used, but this time considering an ultralight boson mass. Then it was shown that not only RCs are acceptable, but also that the model was a serious model at galactic core scales as well. Unfortunately, it was shown that such halos are unstable, with decay  time scales of the order of megayears, which is an insufficient lifetime for a halo \cite{Bosonicos2013}; furthermore this model, being a particular set of parameters of that in the TF limit model for the ultralight particle, it inherits the overdetermined set of parameters $R,~a,~m$ of the original model in \cite{BohemerHarko2007}.

Summarizing the status of spherically symmetric BEC halo models with bosons in the same ground state: i) the ground state solutions of the GPP system have good properties of stability and are late-time attractor solutions \cite{GuzmanUrena2004,GuzmanUrena2006,BernalGuzman2006a}, however show an unacceptable density profile, and ii) the halos constructed in the TF limit \cite{BohemerHarko2007,RoblesMatos2012} have an overdetermined set of parameters and are unstable, and thus are not considered acceptable anymore. Therefore the model needs to be constructed from the beginning at local scales and explore possible solutions.

One would like to preserve the ultralight nature of bosonic dark matter that works fine at cosmic scales, construct galactic halos that are stable and preferably late-time attractors in time, with appropriate core density, and with a set of parameters that allows the existence of different halo sizes, avoiding the unphysical overdetermination of parameters (in the case of the TF limit spherical model). We thus  explore the possibility that a local parameter, possibly different for each galaxy can help. A good such parameter is angular momentum. Adding angular momentum to a BEC fluctuation may improve the RCs in galaxies, explain the different sizes of galaxies, all this retaining the good properties of the model at cosmic scales. 

There are some precedents dealing with rotating BEC dark matter halos, for instance in \cite{RindlerShapiro2012}, spheroid and ellipsoid analytic solutions to the GPP system with rotation are studied as rotating BEC dark matter halos in various scenarios, and the results particularly focus on the possibility of vortex formation. However these analytic results are limited by the conditions imposed to obtain exact solutions, particularly regimes of domination (self-interaction and quantum pressure terms). In order to have a more general picture it is necessary to solve the GPP system with rotation under very general conditions. The challenge presented in this way involves the solution of the GPP system out of spherical symmetry, and in order to study general scenarios it is necessary to solve the GPP system numerically. In this sense, some numerical implementations designed to solve the GPP system in 3D have been presented with different purposes, for instance \cite{GuentherThesis,Choi,Wai,Madarassy}. 

In this paper we firstly present our numerical code that solves the GPP system in 3D, with various tests such a code must satisfy. Secondly, as a first application of the code, we show the effects of applying angular momentum to spherically symmetric configurations made of an ultralight boson, and particularly focus on the spatial redistribution of bosons, and its direct implication on galactic RCs.

In the following section we describe the numerical methods used to solve (\ref{eq:SPcompleta}) for a rotating configuration in 3D. In section \ref{sec:RCs} we present the RCs for rotating configurations and in section \ref{sec:conclusions} we draw some conclusions.


\section{Numerical Methods}
\label{sec:numerics}

We solve system (\ref{eq:SPcompleta}) numerically in cartesian coordinates.
It is a coupled system consisting of an evolution equation for $\Psi$ with a potential that is solution of Poisson equation sourced by $|\Psi|^2$.

The first step before integrating (\ref{eq:SPcompleta}) requires the remotion of constants using the following change of variables $\hat{\Psi} = \frac{\sqrt{4\pi G}\hbar}{mc^2}\tilde{\Psi}$,
$\hat{x} = \frac{mc}{\hbar}\tilde{x}$,
$\hat{y} = \frac{mc}{\hbar}\tilde{y}$,
$\hat{z} = \frac{mc}{\hbar}\tilde{z}$,
$\hat{t} = \frac{mc^2}{\hbar}\tilde{t}$,
$\hat{V} = \frac{\tilde{V}}{mc^2}$,
$\hat{a} \rightarrow \frac{c^2}{2mG}\tilde{a}$, so that the numerical coefficients $\hbar,~\hbar^2/m,~2\pi \hbar^2/m^2,~4\pi Gm$ do not appear in (\ref{eq:SPcompleta}). Additionally, we set our code units allowing accurate calculations, using the invariance of system (\ref{eq:SPcompleta}) under the transformation 
$t = \lambda^2 \hat{t}$, 
$x = \lambda\hat{x}$, 
$y = \lambda\hat{y}$, 
$z = \lambda\hat{z}$, 
$ \Psi = \hat{\Psi}/\lambda^2$,
$V = \hat{V} / \lambda^2$,
$a = \lambda^2 \hat{a}$, 
 for an arbitrary value of the parameter $\lambda$ \cite{GuzmanUrena2004}. These rescaling reduces the original system (\ref{eq:SPcompleta}) to the following one
 
\begin{eqnarray}
i \frac{\partial \Psi}{\partial t} &=& -\frac{1}{2}\nabla^2 \Psi + V\Psi +a|\Psi|^2\Psi, \label{eq:SchroNoUnits}\\
\nabla^2 V &=& |\Psi|^2. \label{eq:PNoUnits}
\end{eqnarray}

\noindent The strategy we follow to solve the coupled GPP system, consists in considering Schr\"odinger equation an evolution equation for $\Psi$, and Poisson equation a constraint that has to be solved every time it is required during the integration of Schr\"odinger equation.

{\it Evolution.} We approximate the GPP (\ref{eq:SchroNoUnits}-\ref{eq:PNoUnits}) system using finite differences on a uniformly discretized grid on a spatial  domain, described with cartesian coordinates. We solve the system on the spatial domain  $[x_{min},x_{max}]\times [y_{min},y_{max}]\times [z_{min},z_{max}]$, using the 3D Fixed Mesh Refinement (FMR) driver Nakode.mx developed by us \cite{Nakode.mx}.

\begin{figure}[htp]
\includegraphics[width=7.5cm]{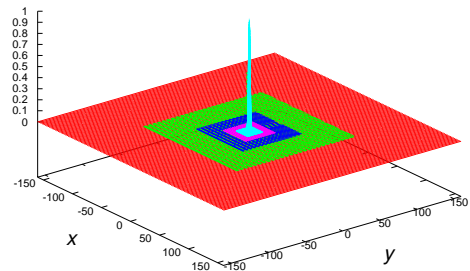}
\caption{\label{fig:FMR} We illustrate the grid structure of our FMR driver with the density of probability at initial time for one of our production runs, projected on the $xy-$plane. We use five refinement levels covering the domain. The different numerical domains covered by each resolution are $[-160,160]^3$, $[-80,80]^3$, $[-40,40]^3$, $[-20,20]^3$ and $[-10,10]^3$. The domain covered with each refinement level uses  $128^3$ points. The highest resolution is kept in the center of the domain, where the density and gravitational potential are steeper.}
\end{figure}

We use FMR with various purposes: firstly because we want to keep a good resolution and therefore numerical accuracy in the central parts of the condensate trap, where the density is mostly concentrated; secondly, because part of the matter will be expelled by rotation and it is important to make sure it is not reflected back into the numerical domain and then contaminate the calculations. In order to cover a sufficiently big numerical domain one would require unaffordable computer memory, however with the use of FMR only high resolution is used in the regions where the functions are steeper, in our case in the center of the condensate.  Our particular set up considers a halved resolution among consecutive refinement levels and boxes in all cases are centered at the coordinate origin, which coincides with the center of the condensate. In this way, our implementation allows high accuracy at the center and the possibility to place the boundaries far away. We illustrate our FMR set up in Fig. \ref{fig:FMR}, where we show $|\Psi|^2$ in a domain for one of our production runs.

For the evolution, Schr\"odinger equation is semi-discretized at each point of the numerical domain, and the values of $\Psi$ evolve using the method of lines, with an iterative Crank-Nicholson time integrator and second order accurate spatial stencils to approximate spatial derivatives. We use the Berger-Oliger evolution algorithm in the FMR, which was adapted to Schr\"odinger equation, which is first order in time and second order in space. 

Considering that the value of $V$ is required during the time integration, Poisson equation is solved for $V$: ($a$) at each of the intermediate steps of the time integration, and ($b$) on each refinement level required, in order to source Schr\"odinger equation for the evolution of $\Psi$.

\begin{figure}[htp]
\includegraphics[width=4.cm]{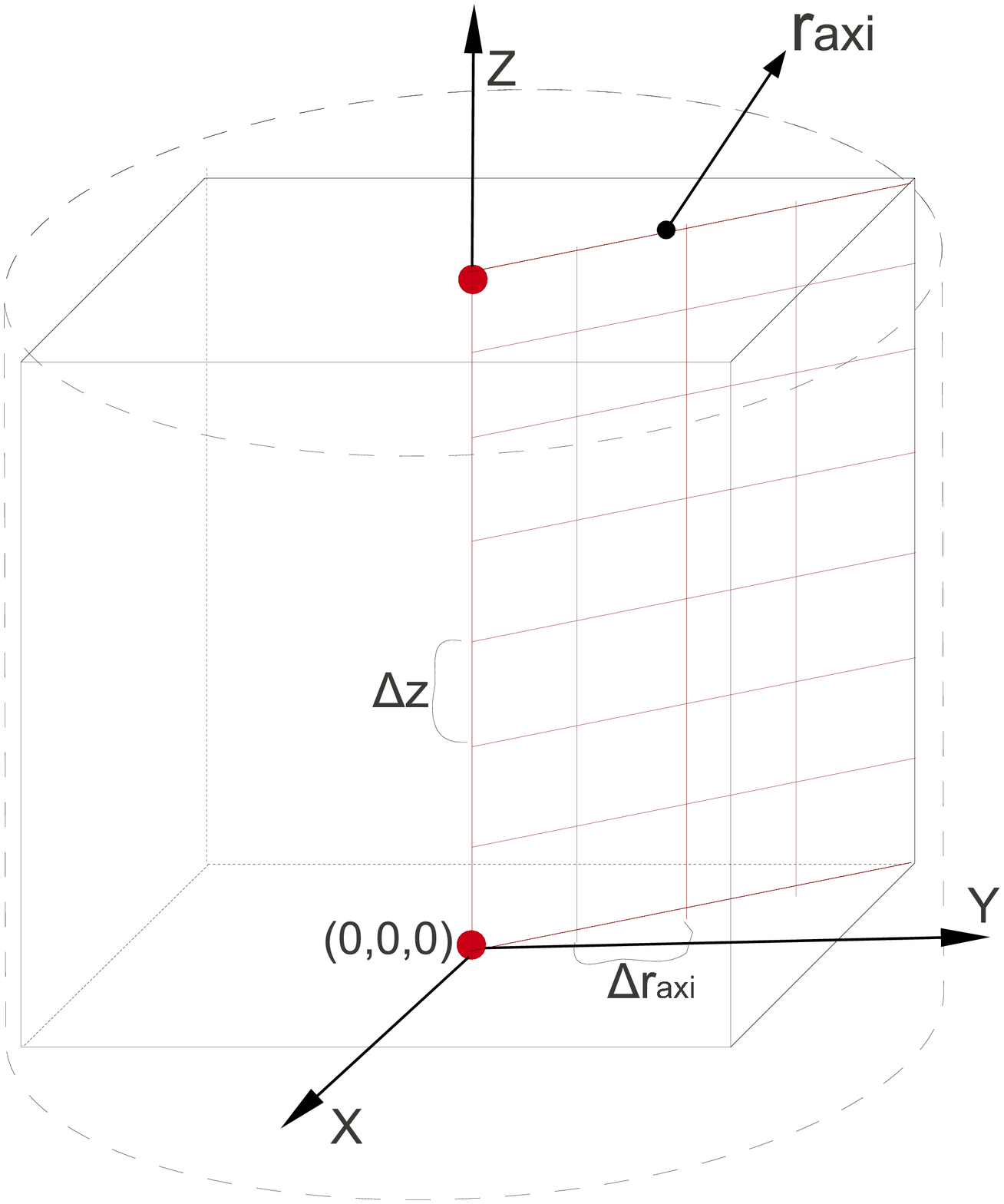}
\caption{\label{fig:axialdomain} We show the diagonal plane orthogonal to $\hat{x}+\hat{y}$, where Poisson equation is defined and solved.}
\end{figure}

{\it Poisson equation.} Solving an elliptic equation every intermediate time step during an evolution problem becomes expensive when there are no symmetries. In order to integrate it efficiently, we integrate Poisson equation only on the diagonal plane across the 3D domain with normal $(\hat{x}+\hat{y})$. This can be done as long as we only analyze axially symmetric configurations. For a given refinement level we show the domain of integration of Poisson equation in Fig. \ref{fig:axialdomain}, defined by the domain $r_{axi} \in [0,\sqrt{x_{max}^2 + y_{max}^2}]\times z \in [z_{min},z_{max}] $ with resolutions $\Delta r_{axi}=\sqrt{\Delta x^2 + \Delta y^2}$ and $\Delta z$. 

The two dimensional Poisson equation is written in cylindrical coordinates, on the diagonal plane, as follows 

\begin{equation}
\frac{\partial^2 V_{axi}}{\partial r^2_{axi}} + \frac{1}{r_{axi}}\frac{\partial V_{axi}}{\partial r_{axi}}+\frac{\partial^2 V_{axi}}{\partial z^2}= |\Psi_{axi}|^2,
\label{eq:pois}
\end{equation}

\noindent where $V_{axi}(r_{axi},z)$ is the unknown potential on the diagonal plane, and $\Psi_{axi}(r_{axi},z)$ is the interpolated value of the wave function from the 3D mesh on this plane at a given time. 

Since the equation is singular along the $z$-axis $r_{axi}=0$, we stagger the plane $\Delta r_{axi}/2$ off  the axis. Additionally,  we avoid the divergence of the second term in (\ref{eq:pois}) substituting $\frac{1}{r_{axi}} \frac{\partial V_{axi}}{\partial r_{axi}} = 2 \frac{\partial V_{axi}}{\partial r_{axi}^{2}}$, where the later term is a derivative with respect to $r_{axi}^{2}$.

The boundary condition on $V_{axi}$ is monopolar $V=-M/\sqrt{x^2+y^2+z^2}$ at the exterior boundary $[r_{axi}=\sqrt{x_{max}^2+y_{max}^2},z]\cup [r_{axi},z_{max}]\cup [r_{axi},z_{min}]$ and a parity condition at the staggered axis $V_{axi}(r_{axi}-\Delta r_{axi}/2,z)=V_{axi}(r_{axi}+\Delta r_{axi}/2,z)$.

Finally, Poisson equation is integrated along the diagonal plane using a Successive Overrelaxation (SOR) algorithm. This solution is calculated for all the refinement levels and at every intermediate time step during the evolution. Each time $V_{axi}$ is interpolated back into the 3D grid, for all the grid points in the domain, based on the fact that the system is axially symmetric. In this way,  $V$ is known in 3D and used as source in Sch\"odinger equation.

{\it Sponge.} It is expected that some of the density of probability will approach the numerical boundary because this is the mechanism of relaxation of the system \cite{SeidelSuen1990,GuzmanUrena2006}. In order to avoid that such excess of matter reflects back into the numerical domain and therefore contaminate the calculations, we implemented a sponge. This consists in the addition of an imaginary potential, thus we redefine the potential in Schr\"odinger equation by adding an imaginary part $V=V+iV_{im}$. The implication of an imaginary potential is that the continuity equation reads $\partial\rho/\partial t + \nabla \cdot [i/2 (\Psi \nabla \Psi^* - \psi^* \nabla \Psi)] = 2 V_{im}|\Psi|^2$, and then, as long as $V_{im}<0$ the imaginary potential acts as a sink of particles.

\begin{figure}[htp]
\includegraphics[width=8cm]{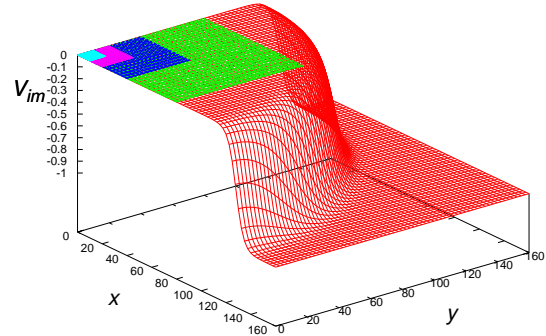}
\caption{\label{fig:esponja} We show a projection of $V_{im}$ on the $xy$-plane. The region where $V_{im}$ is non-zero is where the sponge captures the density of probability approaching the boundary. The numerical parameters used in the production runs are $V_0 = 1$, $\delta=8$ and $r_c =128$.}
\end{figure}

The specific profile chosen for $V_{im}$ is $V_{im}= -\frac{V_0}{2}[2+\tanh(r-r_c)/\delta - \tanh(r_c/\delta)]$, which is a smooth version of a step function, where $V_0$ is the amplitude of the imaginary potential, $r_c$ is the location and $\delta$ is the width of the step. This profile has been used successfully in the past for spherical and axisymmetric cases in 2D  \cite{GuzmanUrena2004,GuzmanUrena2006,BernalGuzman2006a,BernalGuzman2006b}. In Fig. \ref{fig:esponja} the sponge used in the production runs is shown; five refinement levels are used and the sponge covers the domain with the coarsest resolution.

{\it Tests.} In order to show the code works properly, we show in the appendix two tests of the code, which include the unitarity of the evolution algorithm for an exact solution of Schr\"odinger equation and the evolution of  ground state equilibrium solutions of the GPP problem in spherical symmetry with our 3D code. This shows the accuracy of both, the evolution algorithm and the Poisson equation solver.

\subsection{Initial data}

In order to study the effects produced by rotation we can choose an initial profile of the wave function (and thus of the initial density profile) with rotation. However, we choose to start up our simulations with a wave function $\Psi(t=0,{\bf x})$, which corresponds to a ground state equilibrium configuration constructed following \cite{GuzmanUrena2004} and then rigid rotation is added to such configuration. This technique has been applied to the cases when linear momentum is added to equilibrium configurations for the study of collision of configurations and orbital motion \cite{Choi,BernalGuzman2006b}. Then, for a spherically symmetric ground state configuration constructed in spherical symmetry following \cite{GuzmanUrena2004}, we have the wave function function $\Psi_0$. Then we apply a rotation 
$\Psi = e^{-i {\bf L}\cdot {\bf \hat{n}} \theta}\Psi_0$.

\noindent In our particular case we choose a rotation around the $z$-axis an amount $\theta = \arctan(y/x)$ with ${\bf L} = L_z \hat{\bf z}$. In order to parametrize the rigid rotation we choose $L_z = x p_y - yp_x$ to be a constant. We then start up the evolution with this wave function.

\subsection{Monitoring the results}

{\it Total energy.} The stability of a configurations depends highly on whether or not a system is bounded. Thus it is important to calculate during the evolution the total energy of the system. Our energy estimate is calculated in terms of the expectation value of the total energy of the system, which has three contributions, kinetic, gravitational and self-interaction energies. The expectation value of these quantities are respectively

\begin{eqnarray}
K &=& -\frac{1}{2}\int \Psi^* \nabla^2 \Psi dxdydz,\nonumber\\
W &=& \frac{1}{2}\int V |\Psi|^2 dx dy dz,\nonumber\\
I &=& a\int |\Psi|^4 dxdydz
\end{eqnarray}

\noindent where the integrals are calculated numerically on the whole numerical domain. Then the total energy of the system is defined by $E=K+W+I$ and can be calculated at every time step..




{\it Mass.} Even though the integral of the density of probability is conserved during the evolution in the whole space, we calculate the solution of the GPP system in a finite domain that allows the particles to get off, or equivalently to be absorbed by a sponge once they are near the artificial boundaries of our numerical domain. The useful mass estimate will be the integral of the density of probability in the  numerical domain

\begin{equation}
M = \int |\Psi|^2 dx dy dz.
\end{equation}

\noindent In all our simulations in this paper we use a domain consisting of cubic boxes of size $[-160kpc,160kpc]^3$. We estimate the mass and the expectation values of the operators described above within a box of $[-40kpc,40kpc]^3$ only, which is our working physical domain. The domain outside this later box serves as a buffer zone that we use to absorb any potential noises in the calculations and contains the sponge as well.

{\it Rotation curve.} In order to estimate the rotation curve, we place various detectors that measure the tangential velocity $v$ of a test particle. The detectors are located at a set of points along the $x$-axis, and at each of such points we estimate the velocity of a test particle. For that we assume the test particle describes a circular orbit, as usually considered for RCs. Thus  the gravitational force due to the dark matter compensates the centrifugal force according to the usual formula $v(r) = \sqrt{2GM(r)/r}$, where $r$ is the distance from the coordinate center to the detector, and $M(r)$ is the mass contained within a sphere of radius $r$. Explicitly, if a particular detector is located at $(x,y,z)=(x_d,0,0)$ we calculate $v(x_d)$ as

\begin{equation}
v^2(x_d) = \frac{2 G}{|x_d|} \int |\Psi|^2 dxdydz,
\end{equation}

\noindent where the volume integral is calculated in a sphere of radius $x_d$ on our 3D cartesian grid. We place a considerable number of these detectors and construct the RC during the evolution. Clearly, the RC profile will strongly depend on $|\Psi|^2$, which is expected to be different for different values of $L_z$.

\section{Evolution of the system}
\label{sec:RCs}

First of all the code units have to be specified. The scale invariance parameter $\lambda$ is fixed using the spatial coordinates such that $\lambda = \frac{\hbar}{mc}\frac{x}{\tilde{x}}$. In order to use a numerical domain with coordinate values in kpc, that is $x=\tilde{x}$, it suffices to write down the factor $\frac{\hbar}{mc}$ in kpc. For $m=10^{-23}$eV$/c^2$ its value is $\lambda = \frac{\hbar}{mc}[kpc]=0.0006399$. We fix this scaling parameter for the cases explored in this paper. With this value one recovers the mass in solar masses $\tilde{M}=\frac{\hbar c}{mG}\lambda M$, the velocity in km/s $\tilde{v} = c \lambda v$, the energy in Joules $\tilde{E}=\frac{\hbar c^3}{4\pi G m}\lambda^3 E$ and the potential that transforms with the same formula.

Our numerical set up is as follows. Initially, all the cases we analyze show a density distributed within a radius smaller than 5kpc. In order to allow the configuration to relax within a bigger domain, we set production runs for cubic numerical domains with $x_{min}=y_{min}=z_{min}=-160$ and $x_{max}=y_{max}=z_{max}=160$. We use five refinement levels covering boxes with domains $[-160,160]^3$, $[-80,80]^3$, $[-40,40]^3$, $[-20,20]^2$ and $[-10,10]^3$ respectively. Even if the sponge helps at absorbing the density of probability that approaches the boundaries, our FMR implementation allows the possibility of pulling the boundaries this far, in order to obtain results as free as possible of numerical errors reflected from the boundaries.

We add angular momentum with rigid rotation to an originally spherically symmetric equilibrium  configuration. These spherical equilibrium configurations have negative total energy and are stable in a very general sense \cite{GuzmanUrena2006,BernalGuzman2006a}. However, the addition of angular momentum changes the energy of the system already at initial time.  We present the evolution of the system for various initial values of $L_z$ and study in each situation the properties of the resulting configuration. We explore the evolution for two values of $a=0,~0.5$ and show the behavior is pretty similar in both cases.

\subsection{Case $a=0$.}

This is the case of the free field, in which the bosons do not interact and immediately implies there is no energy of self-interaction $I=0$. We present results for the angular momentum $L_z = 0.6,~ 0.85, ~0.95$ which are representative of three different responses of the system. 

The case with $L_z=0.6$, starts with negative total energy and therefore it is not expected that the angular momentum distorts significantly the configuration. In the two cases with higher angular momentum, the effect of the addition of angular momentum increases the total energy of the system which results to be positive $E=K+W>0$, which in turn implies that the configurations are initially gravitationally unbounded, however in one case the system relaxes and becomes bounded. Explicitly, the three different cases show the following properties.

\begin{figure}[htp]
\includegraphics[width=4cm]{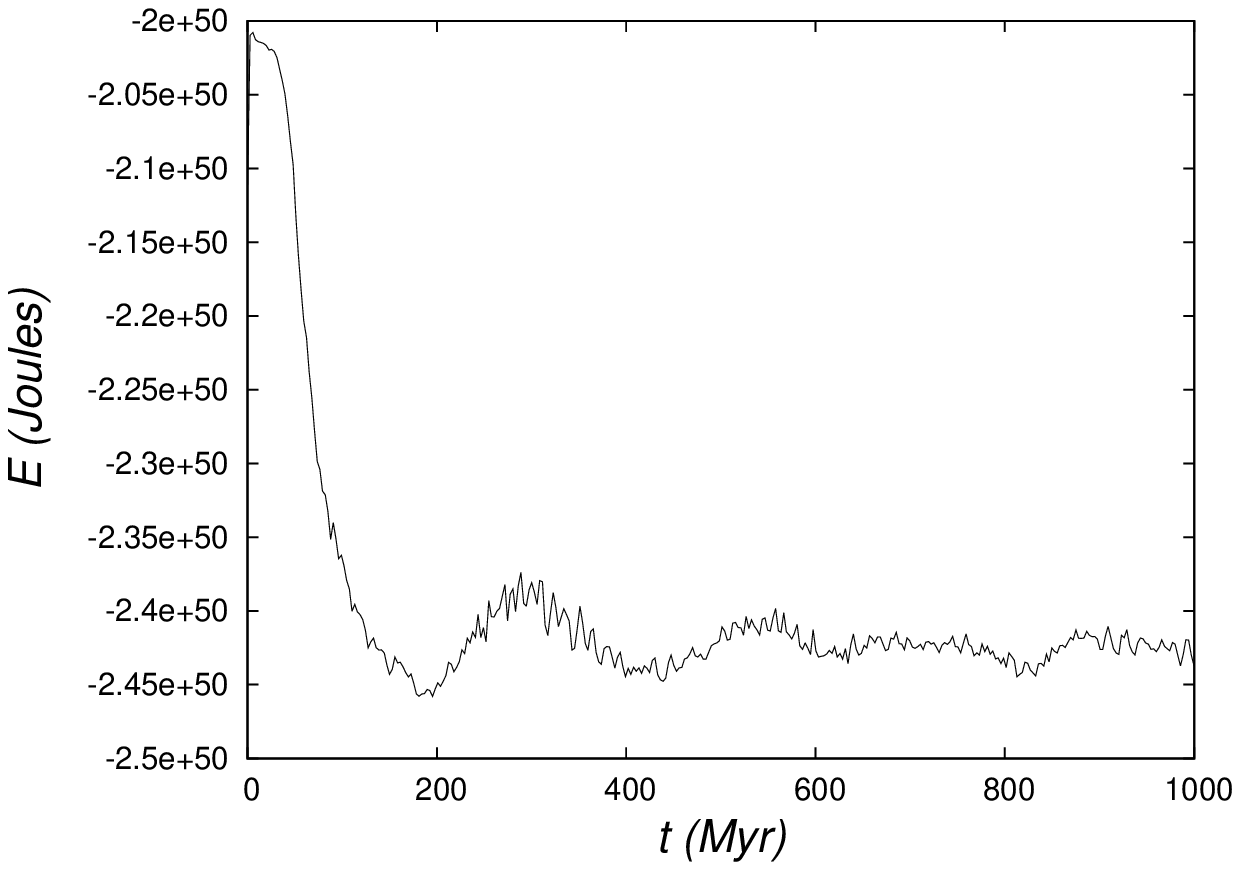}
\includegraphics[width=4cm]{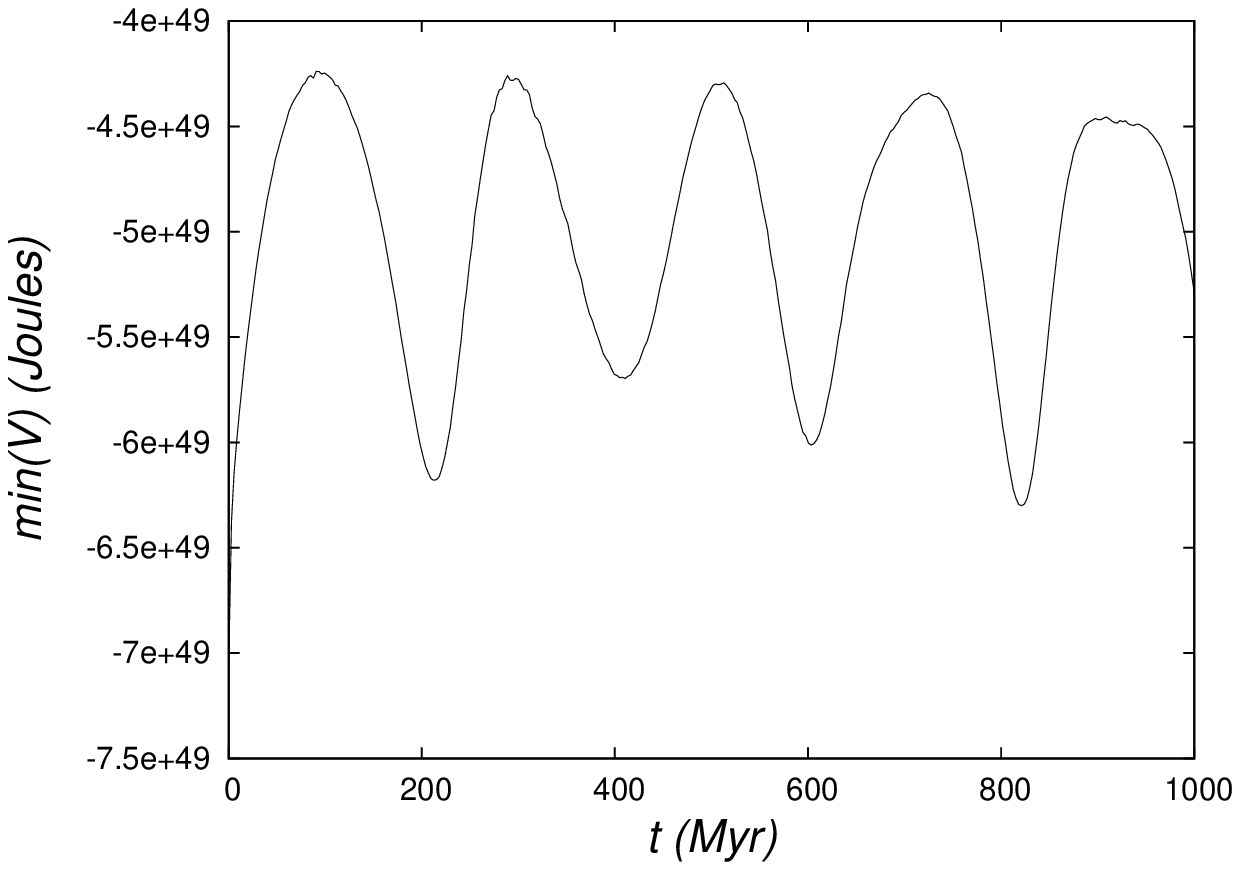}
\includegraphics[width=4cm]{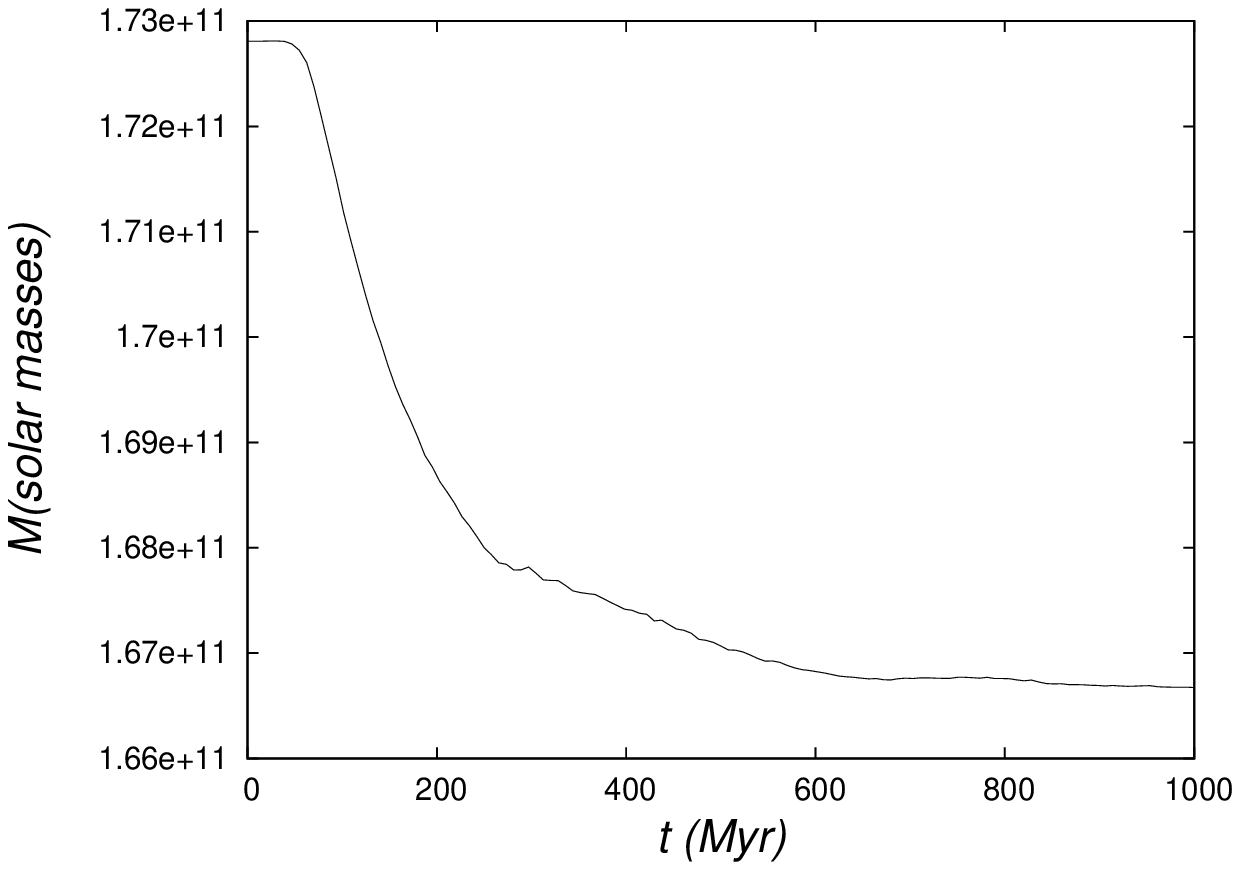}
\includegraphics[width=4cm]{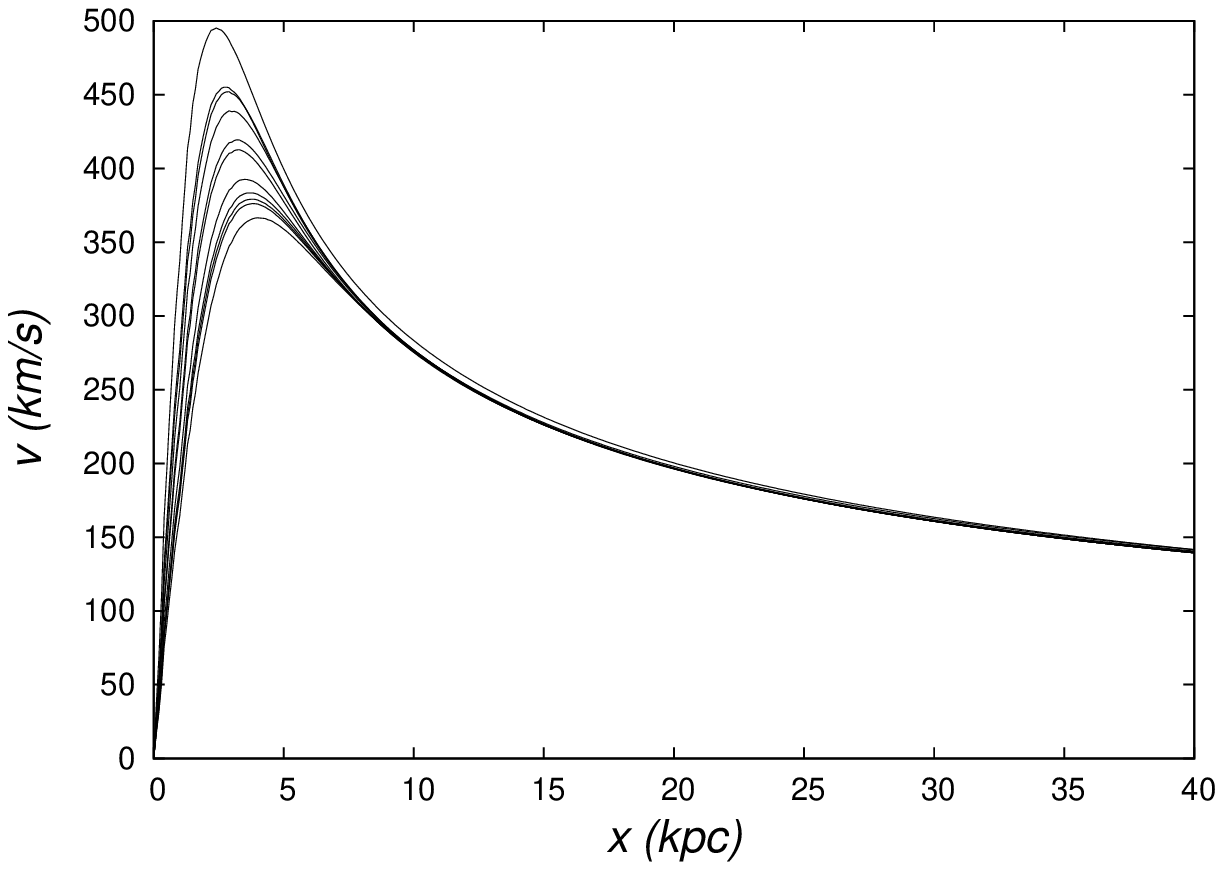}
\caption{\label{fig:Lz0.6} We show the properties of the configuration for $L_z=0.6$. The total energy $E$ remains negative all the way. The minimum of the potential oscillates, which is an indication that the bosons remain confined to a small region and as a consequence, the RC starts and remains  Keplerian after a few kpc. The mass stabilizes, which is an indication that the system is relaxing. The behavior is pretty much that of a perturbed ground state configuration.}
\end{figure}

{\it Case $L_z=0.6$.} In this case the initial total energy is negative and the angular momentum applied to the spherical configuration does not suffice to redistribute the density of bosons in such a way that we observe flat RCs. Instead, we notice in Fig. \ref{fig:Lz0.6} that the gravitational potential dominates over the kinetic energy, confines the configuration and $E<0$ all the way. The RCs are Keplerain at scales of the order of a few kpc, as happens for the spherically symmetric ground state solutions of the GPP system.

\begin{figure}[htp]
\includegraphics[width=4cm]{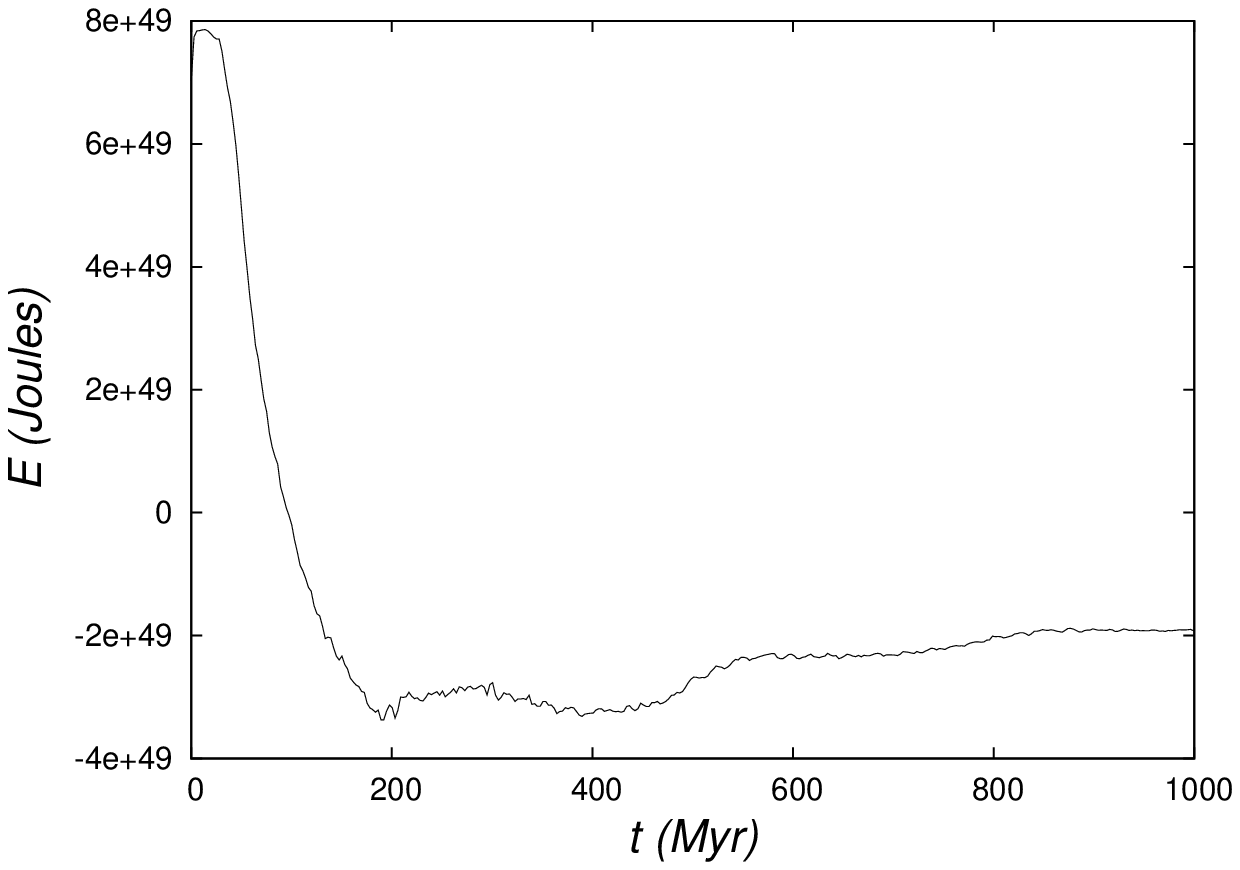}
\includegraphics[width=4cm]{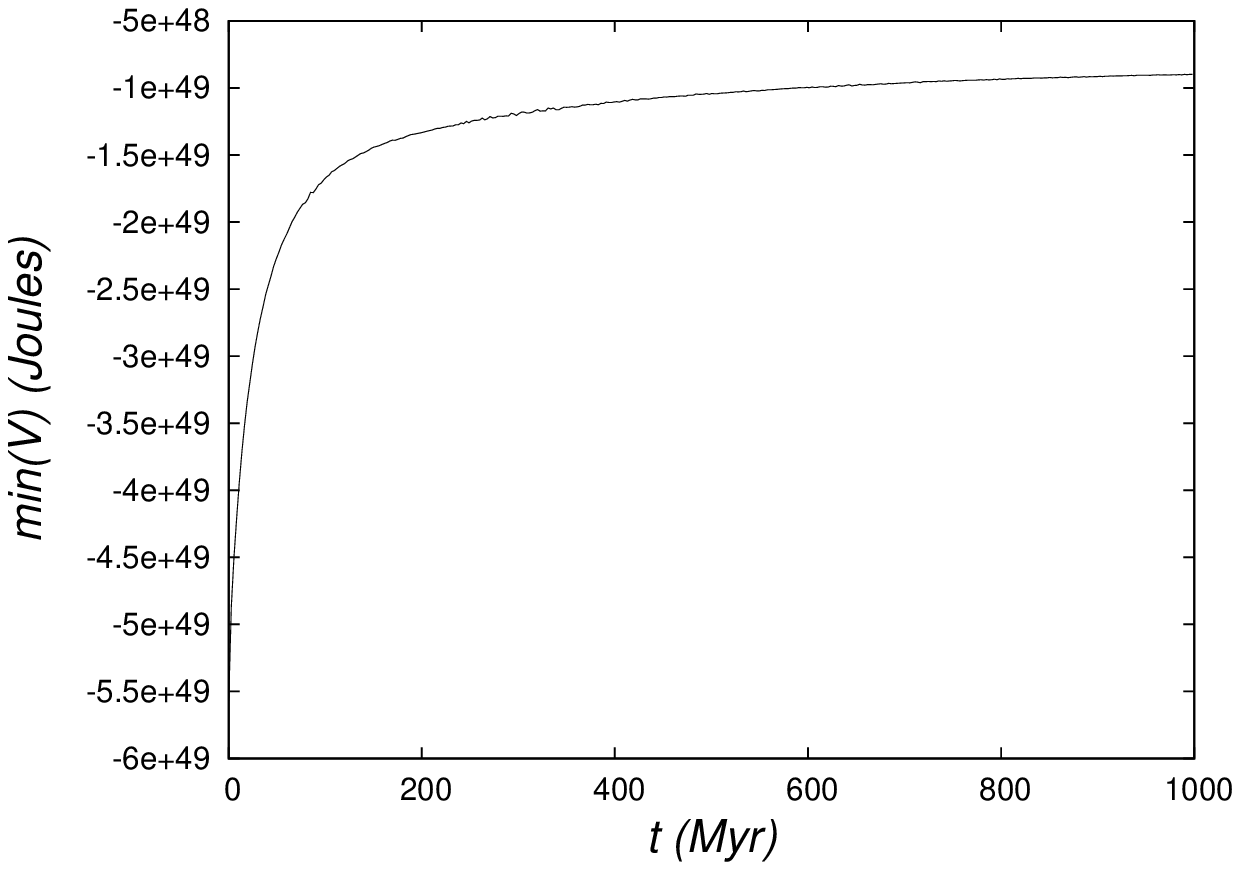}
\includegraphics[width=4cm]{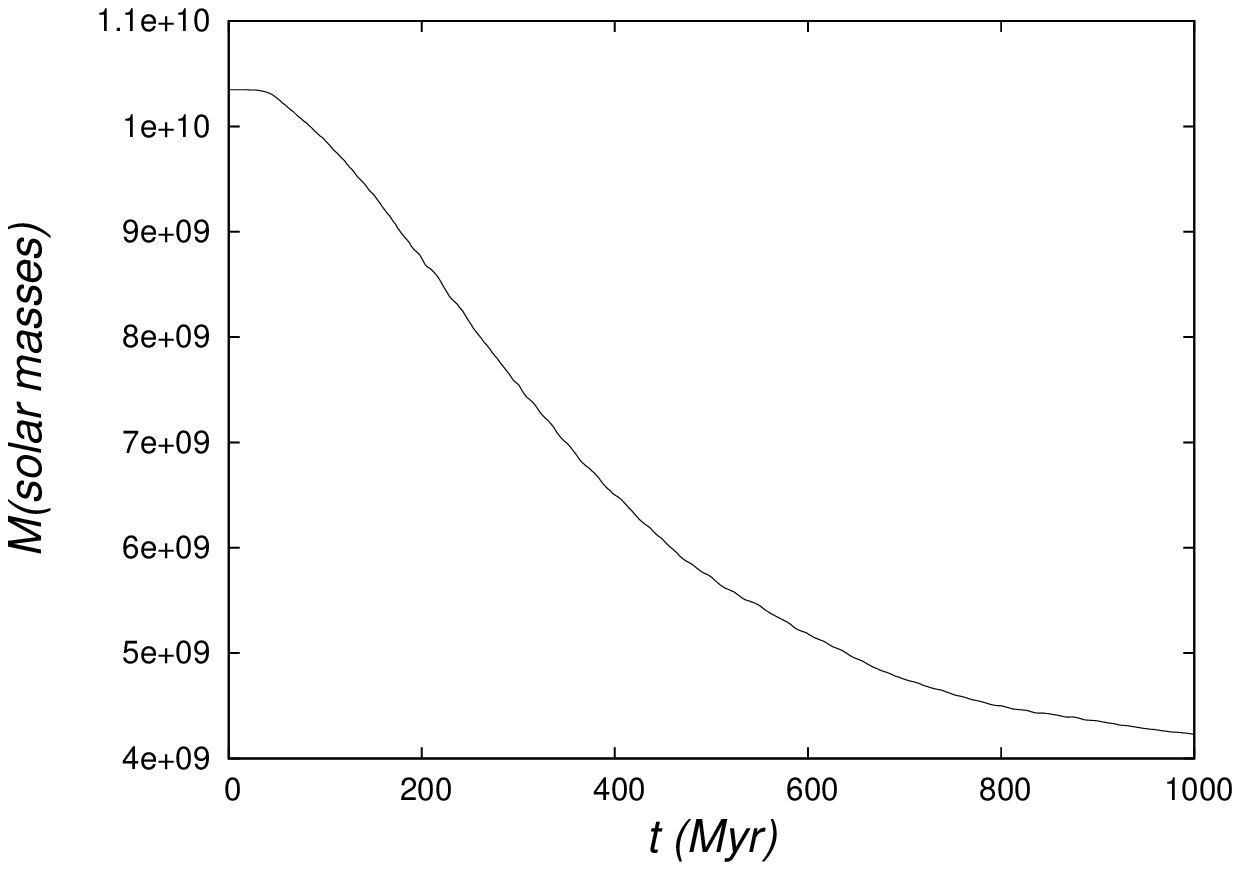}
\includegraphics[width=4cm]{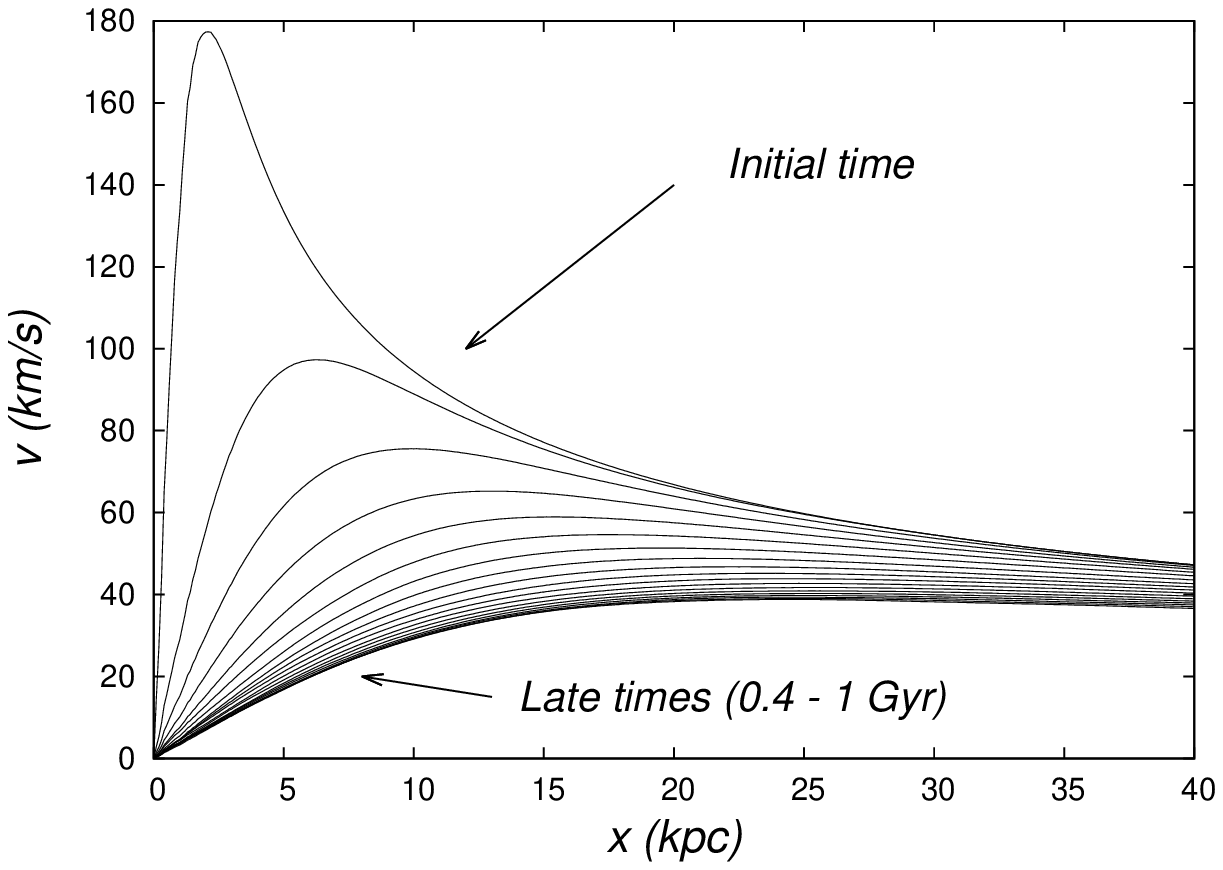}
\caption{\label{fig:Lz0.85} We show the properties of the configuration for $L_z=0.85$. The mass stabilizes around the value of a dwarf galaxy, the energy stabilizes around a negative value and the central potential as well. The rotation curve is Keplerian initially and once the bosons are dispersed away by the rotation, it stabilizes with a shape corresponding to a typical one of a dark matter dominated galaxy.}
\end{figure}

{\it Case $L_z=0.85$.} In this case we observe the desired properties of the configuration that are shown in Fig. \ref{fig:Lz0.85}. First we notice the change of sign of the total energy and how it stabilizes around a negative value, which indicates that the system tends toward a bounded state within the scale of 1Gyr.  Notice also that the gravitational potential stabilizes. The mass approaches a stable vlaue $M \sim 4 \times 10^{10} M_{\odot}$. Finally the RC approaches a very desirable shape, and stabilizes at least during time scales of the order of 1Gyr. With these initial values chosen for the equilibrium configuration, the final nearly equilibrium properties of the system approximate those of a dwarf galaxy.
 
\begin{figure}[htp]
\includegraphics[width=4cm]{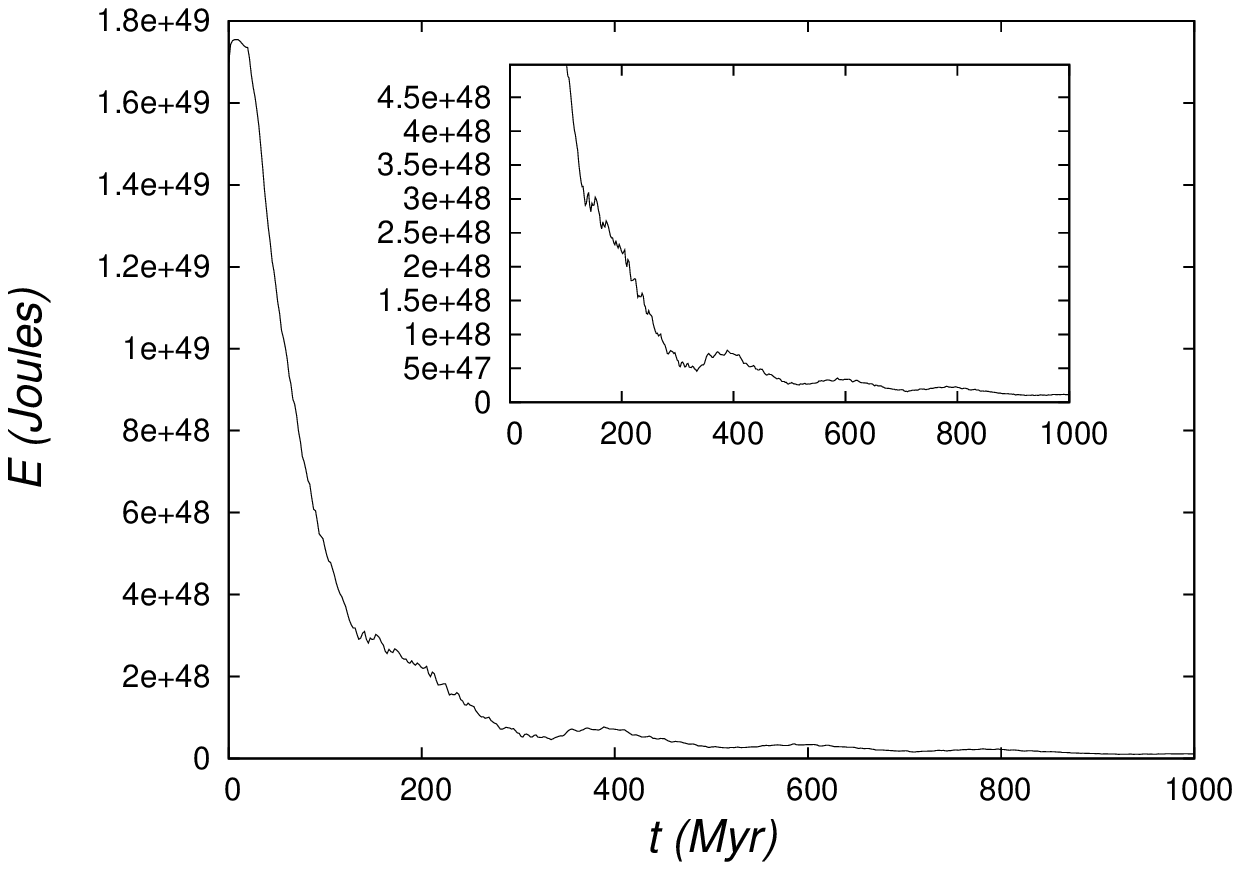}
\includegraphics[width=4cm]{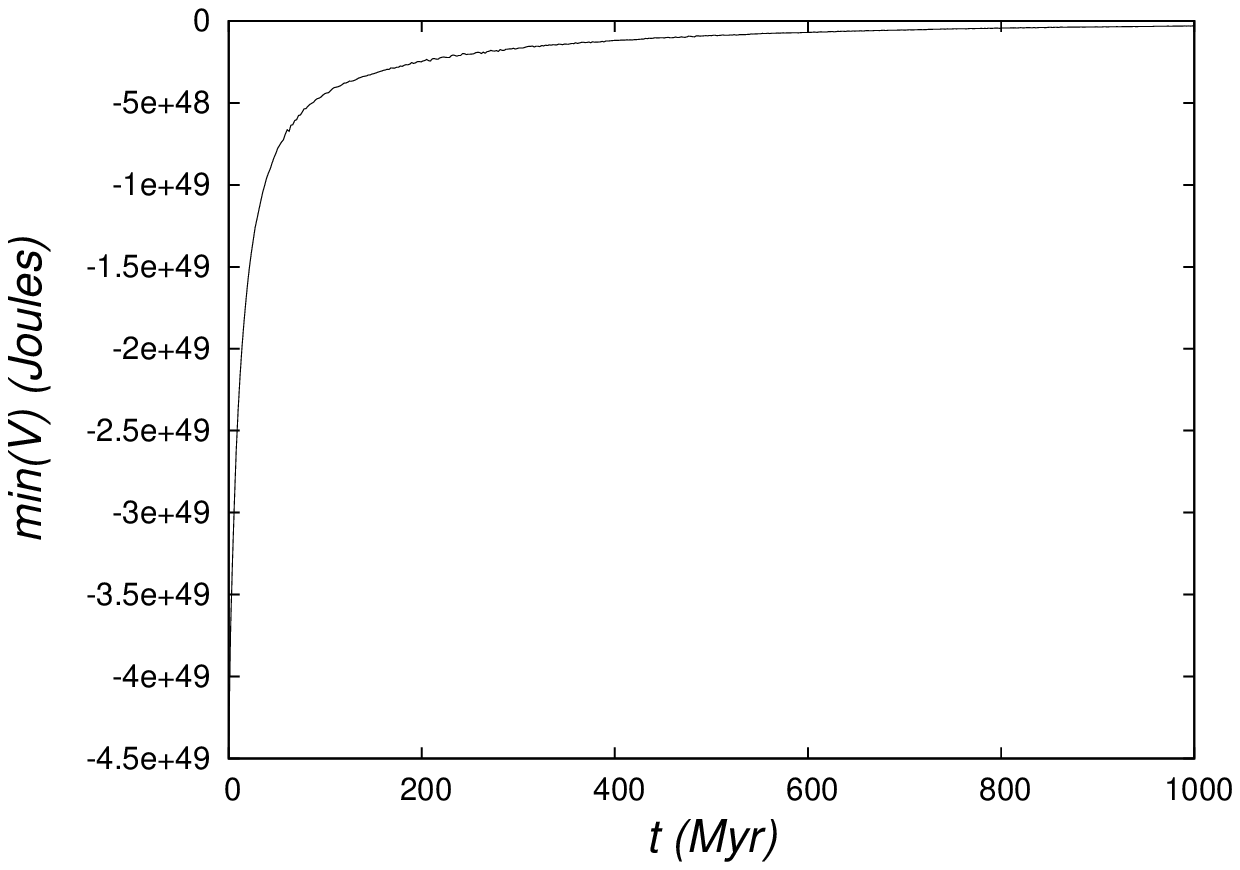}
\includegraphics[width=4cm]{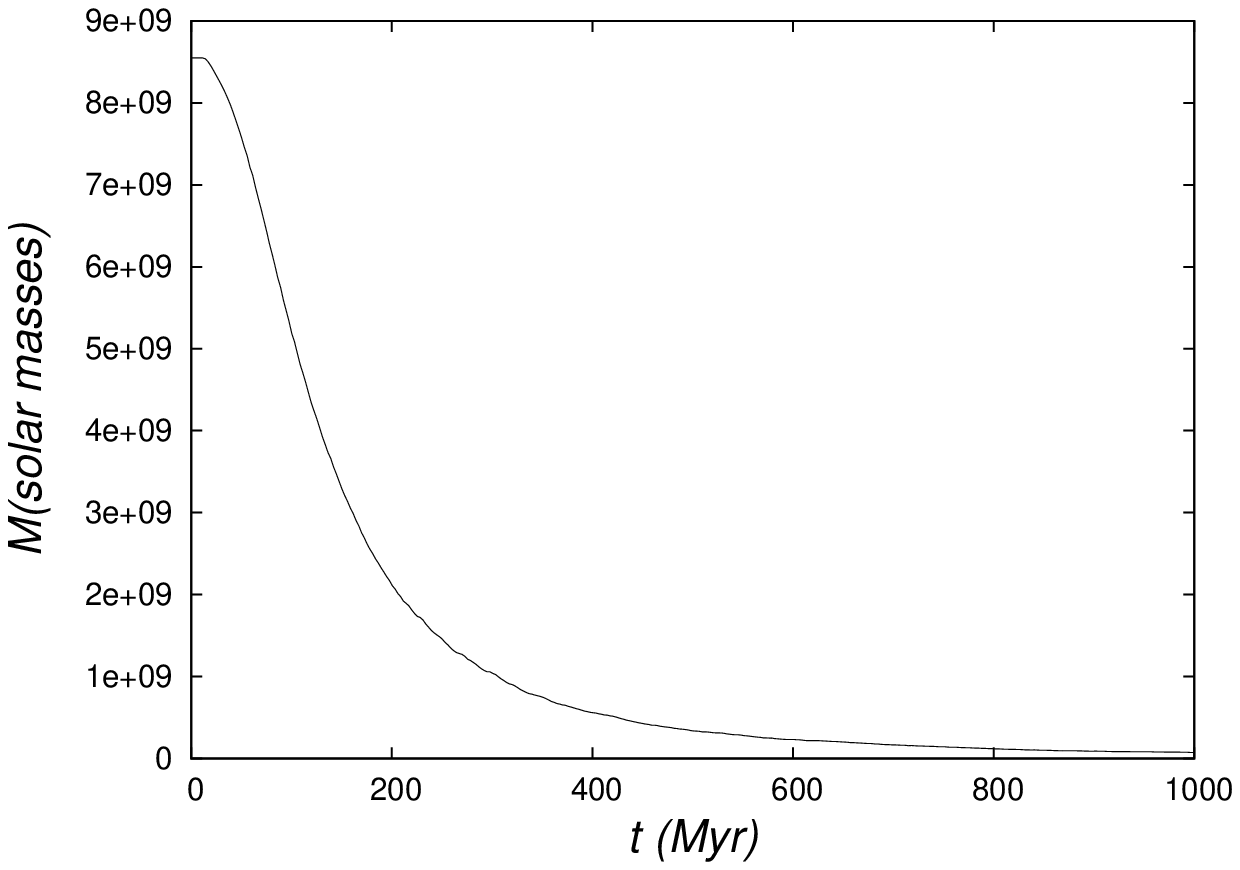}
\includegraphics[width=4cm]{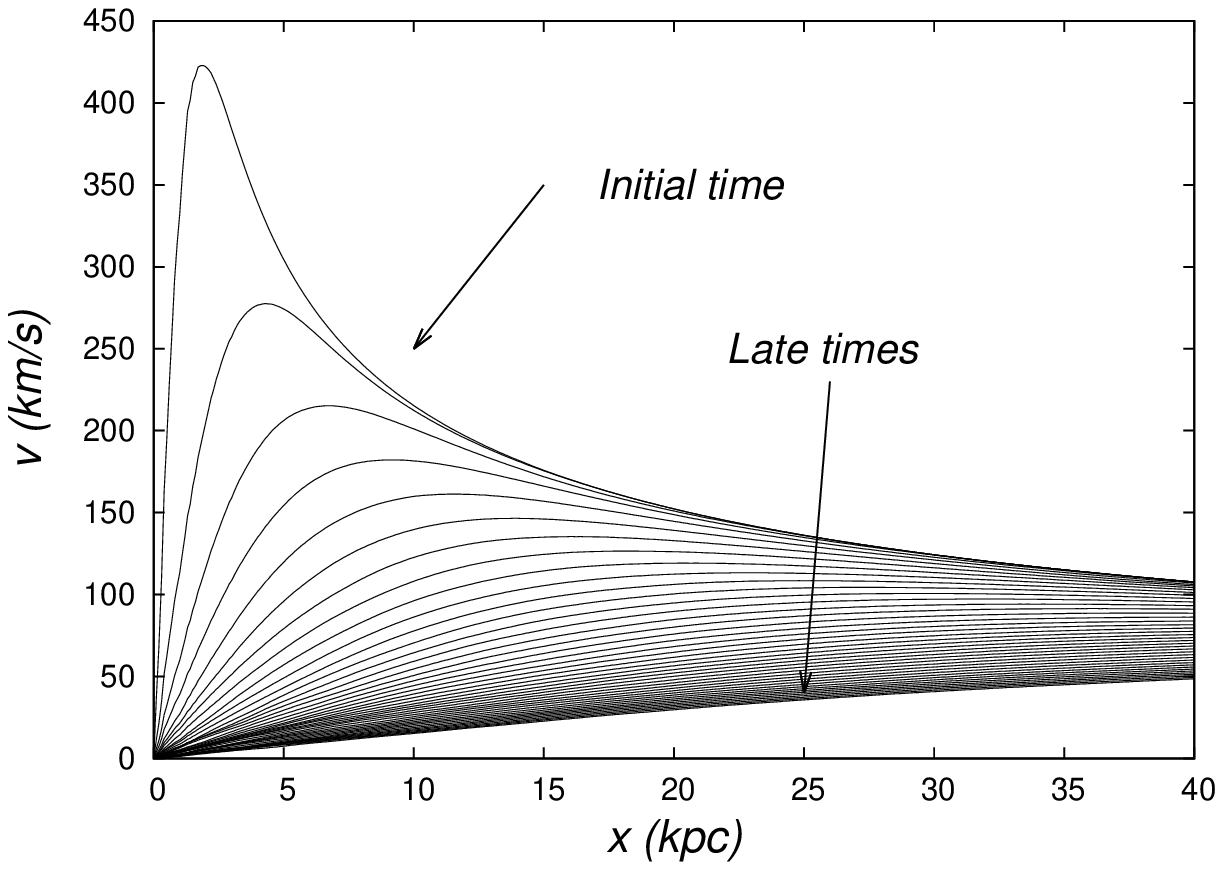}
\caption{\label{fig:Lz0.95} We show the properties of the configuration for $L_z=0.95$. The total energy remains positive, which is a typical indication of an unbounded system. The increase of the gravitational potential and the decrease of the mass within the domain, indicate the matter is being washed out by rotation.}
\end{figure}

{\it Case $L_z=0.95.$} In  this case the angular momentum is so high that redistributes the density of bosons so quickly that they abandon the numerical domain and the configuration is diluted quickly. The physical properties of the configuration are shown in Fig. \ref{fig:Lz0.95}.

\subsection{Case $a=0.5$.}

\begin{figure}[htp]
\includegraphics[width=4cm]{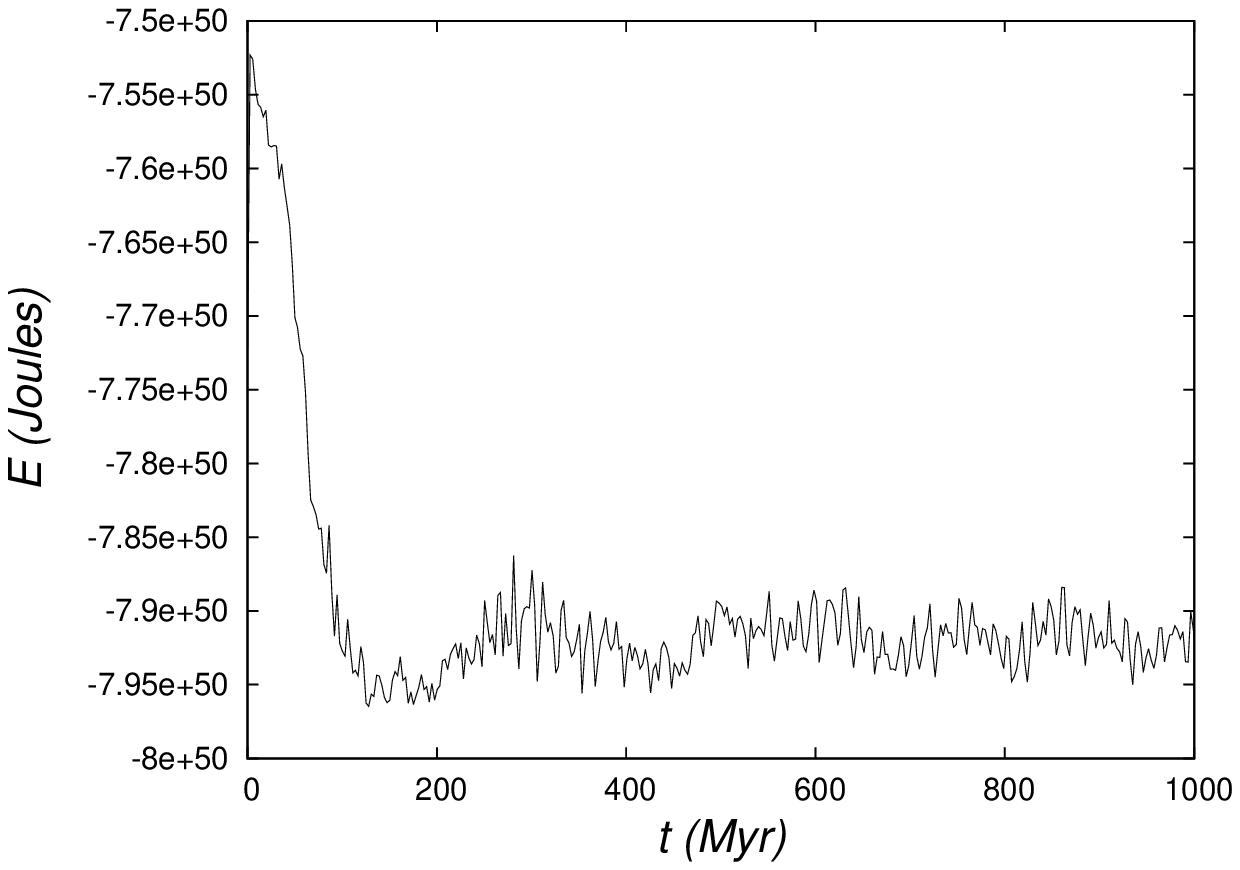}
\includegraphics[width=4cm]{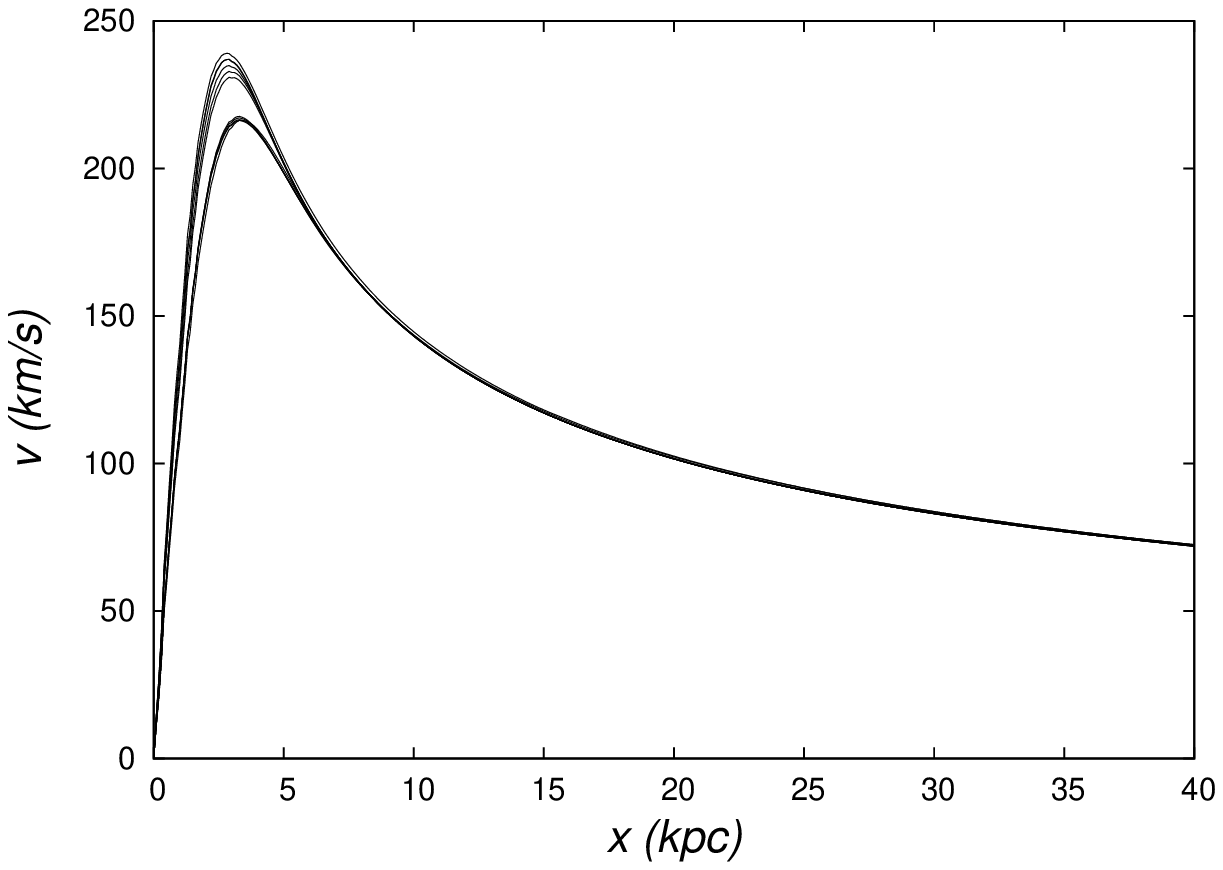}
\includegraphics[width=4cm]{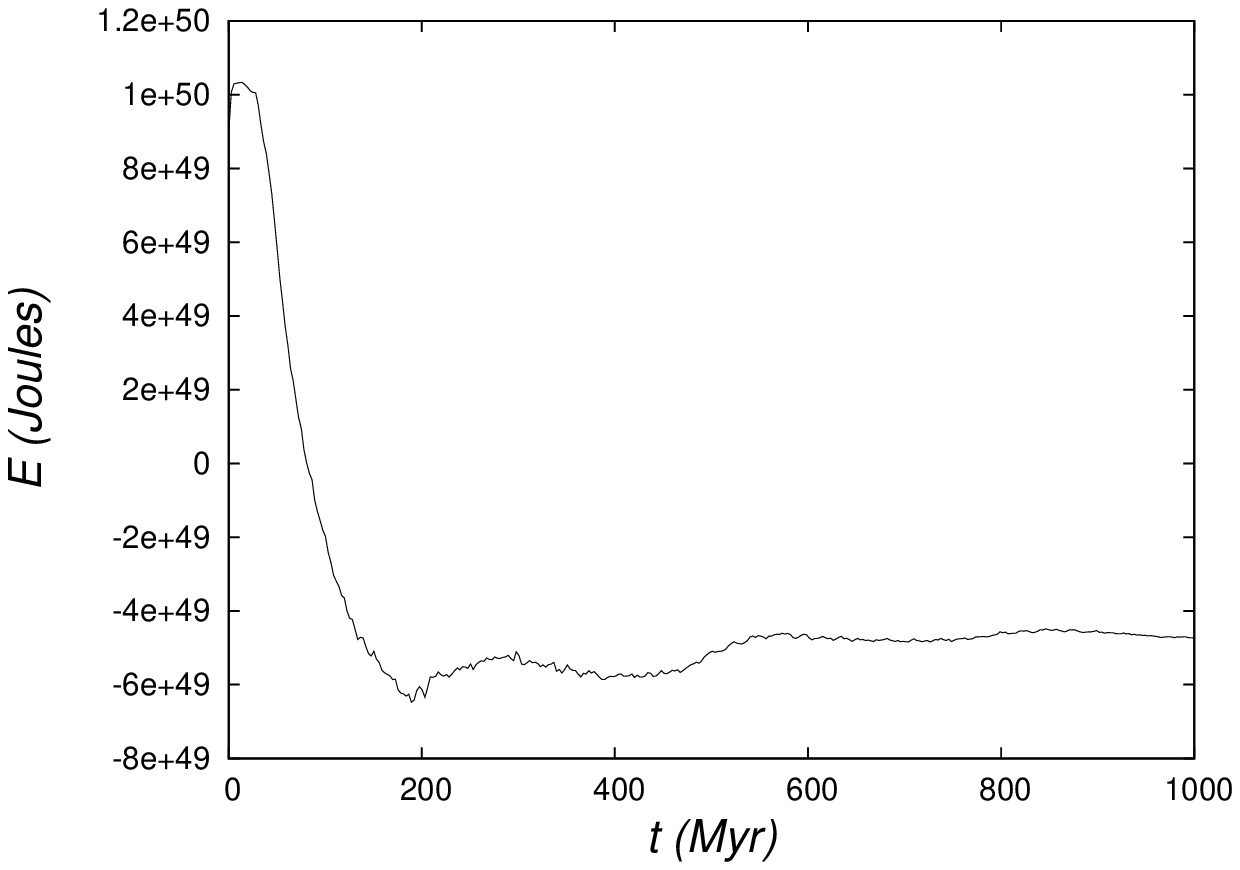}
\includegraphics[width=4cm]{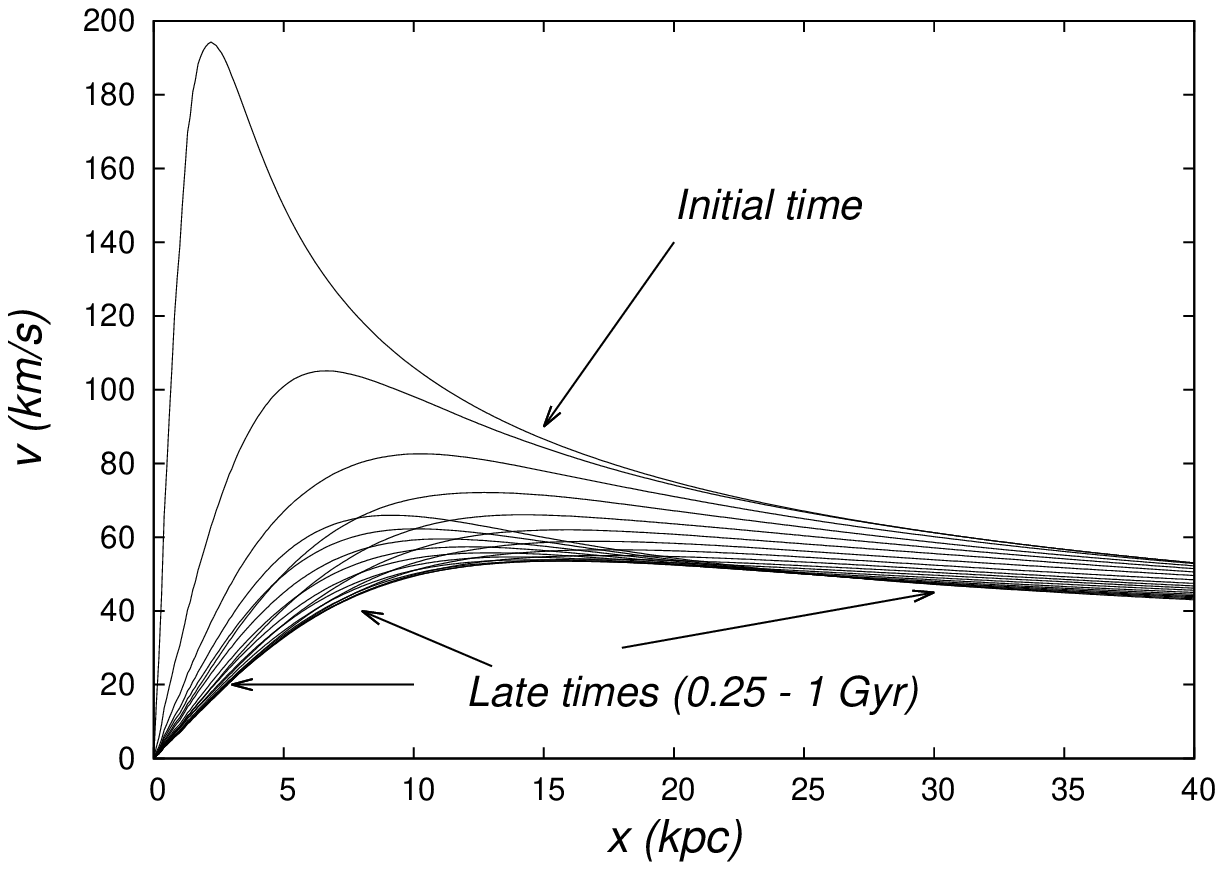}
\includegraphics[width=4cm]{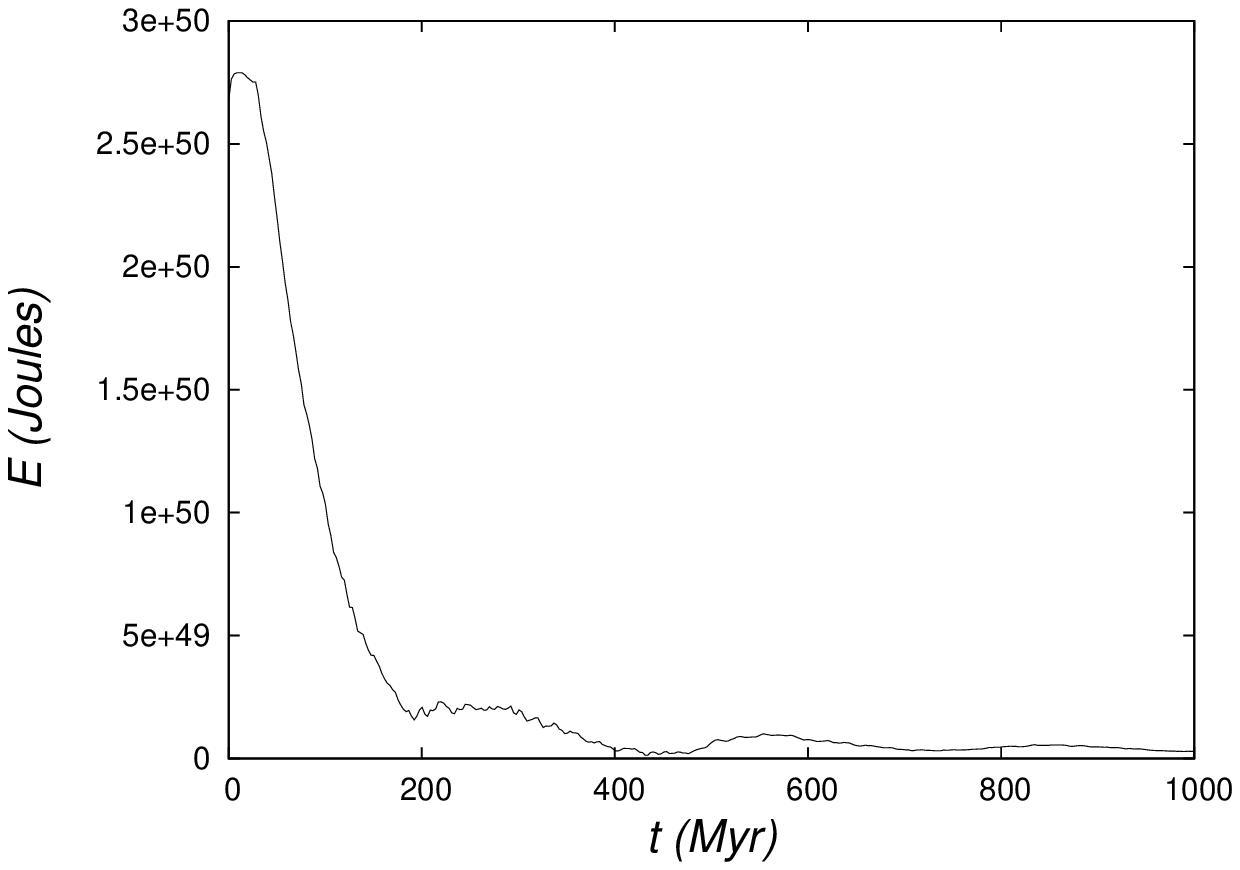}
\includegraphics[width=4cm]{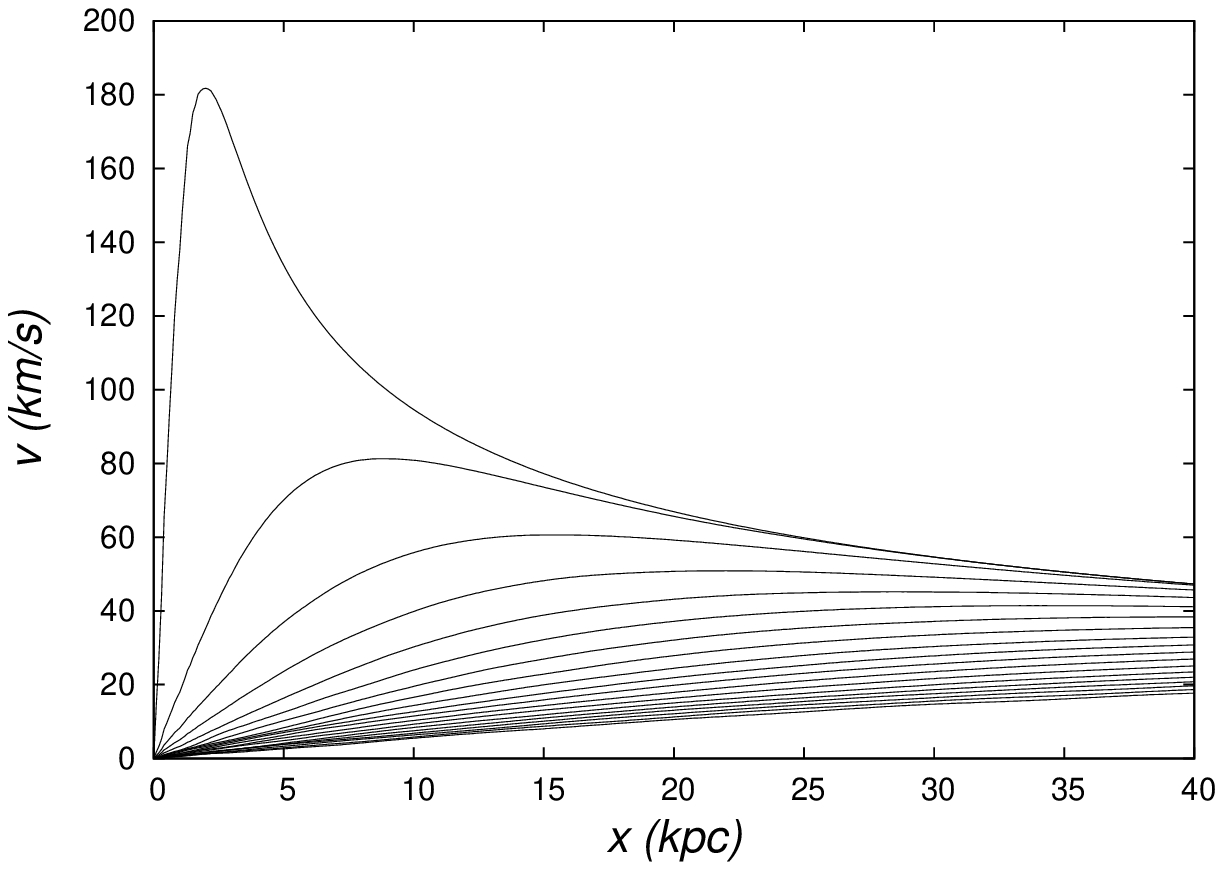}
\caption{\label{fig:a0.5} We show the properties of the configuration for $a=0.5$. In the top panel we show the results for $L_z=0.5$, in which the total energy $E$ remains negative all the way and the RC is Keplerian after a few kpc. In the middle we show the results for $L_z=0.9$ that shows desirable results for dark matter dominated galaxies. finally in the bottom we show the results for $L_z=1$, that corresponds to a configuration washed out by rotation.}
\end{figure}

The addition of a self-interaction term contributes with a repulsive term between bosons and changes the interplay between the kinetic and gravitational terms. This effect is already known in ground state spherically symmetric configurations,  in which more mass can be allocated when the self-interaction is stronger \cite{GuzmanUrena2006}. In this case we show that the same three regimes are also possible when repulsive  self-interaction ($a=0.5>0$) is considered. The three values of the angular momentum we use to represent the three regimes are $L_z=0.5,~0.9,~1$. Since the results are pretty similar to those for the free field case, we present only the total energy and rotation curve in Fig. \ref{fig:a0.5}, for the three cases.

\section{Discussion and Conclusions} 
 \label{sec:conclusions}

We explored the effects of rigid rotation of structures formed by Bose Condensates described by the GPP system of equations. 

There are three regimes: one with  an angular momentum insufficient to disperse the distribution of bosons and provide galactic RCs  in a dark matter dominated galaxy, in the range of order 5 - 20 kpc; a second regime that redistributes the density of bosons in such a way that RCs are pretty comparable to galactic RCs in the range of $\sim 10$kpc, where the configurations stabilize and remain bounded; a third regime in which the rotation is so strong that destroys the configuration and disperses the bosons out of the numerical domain. The life time criterion of our results is given by the time window of 1Gyr we used in all cases.

Knowing that BEC spherically symmetric halo models are poor at explaining galactic dynamics \cite{Bosonicos2013}, what we have shown in this paper is that the model for the ultralight spinless boson DM candidate can work at local scales with the addition of angular momentum to halos. In this way, configurations at local scale are parametrized by the boson mass, the self-interaction of the Bose gas and the angular momentum applied to the halo.

We have also presented a new code that solves the GPP system of equations in 3D . The tests shown are exhaustive and validate this new numerical tool that  will serve for the exploration of the parameter space of rotating BEC halos, including the interplay of self-interaction and rotation, the search of attractor solutions and the study of vortices in a self-gravitating BEC.


\section*{Acknowledgments}

We thank L. A. Ure\~na-L\'opez for valuable comments and suggestions on this paper. 
This research is partly supported by grants
CIC-UMSNH-4.9 and CONACyT 106466. 


\section{Appendix: tests of the code}

\subsection{Unitarity of the algorithm}

\begin{figure}[htp]
\includegraphics[width=4.cm]{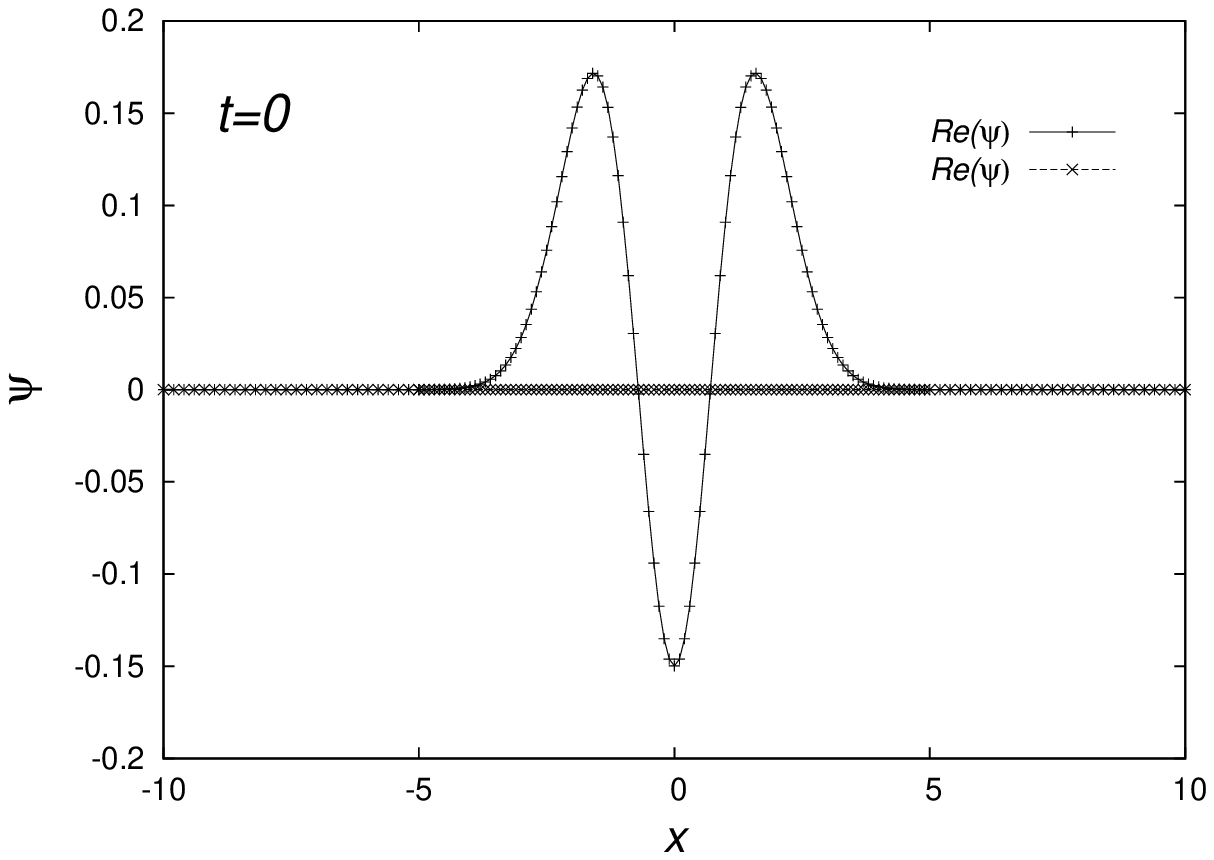}
\includegraphics[width=4.cm]{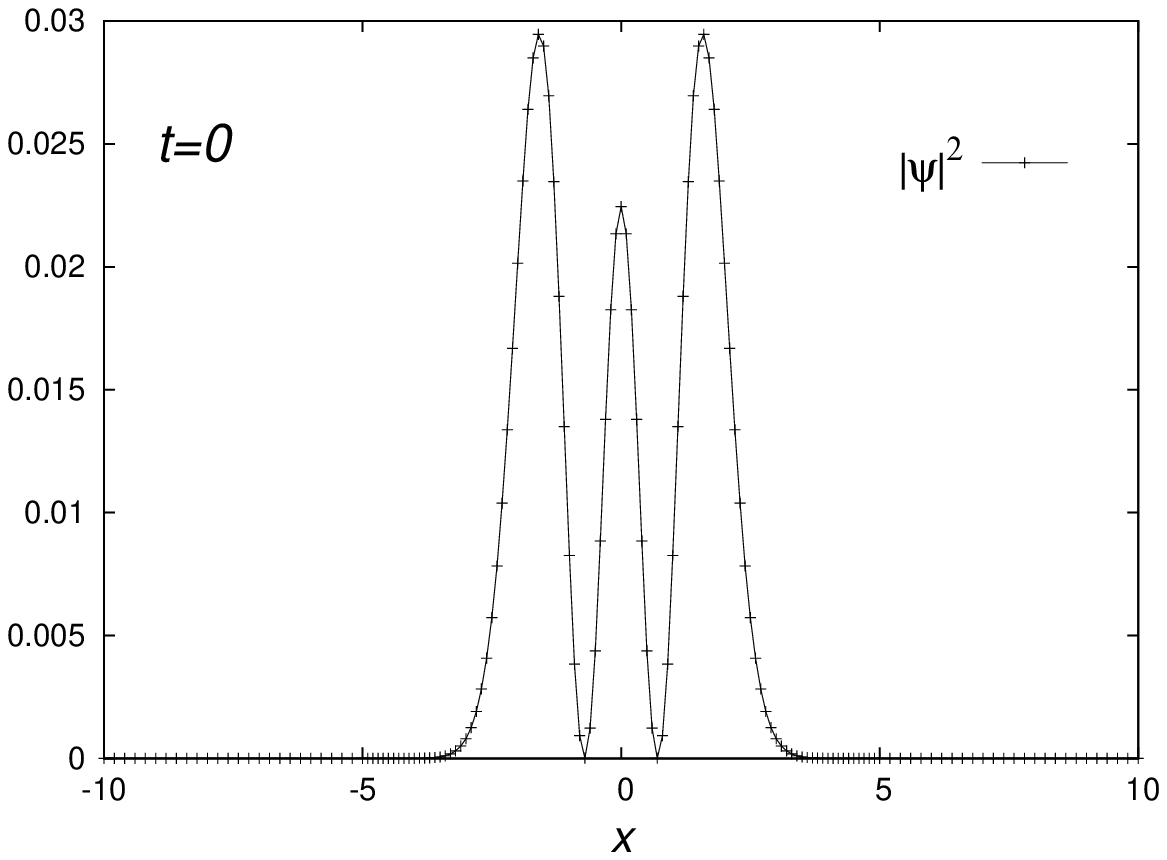}
\includegraphics[width=4.cm]{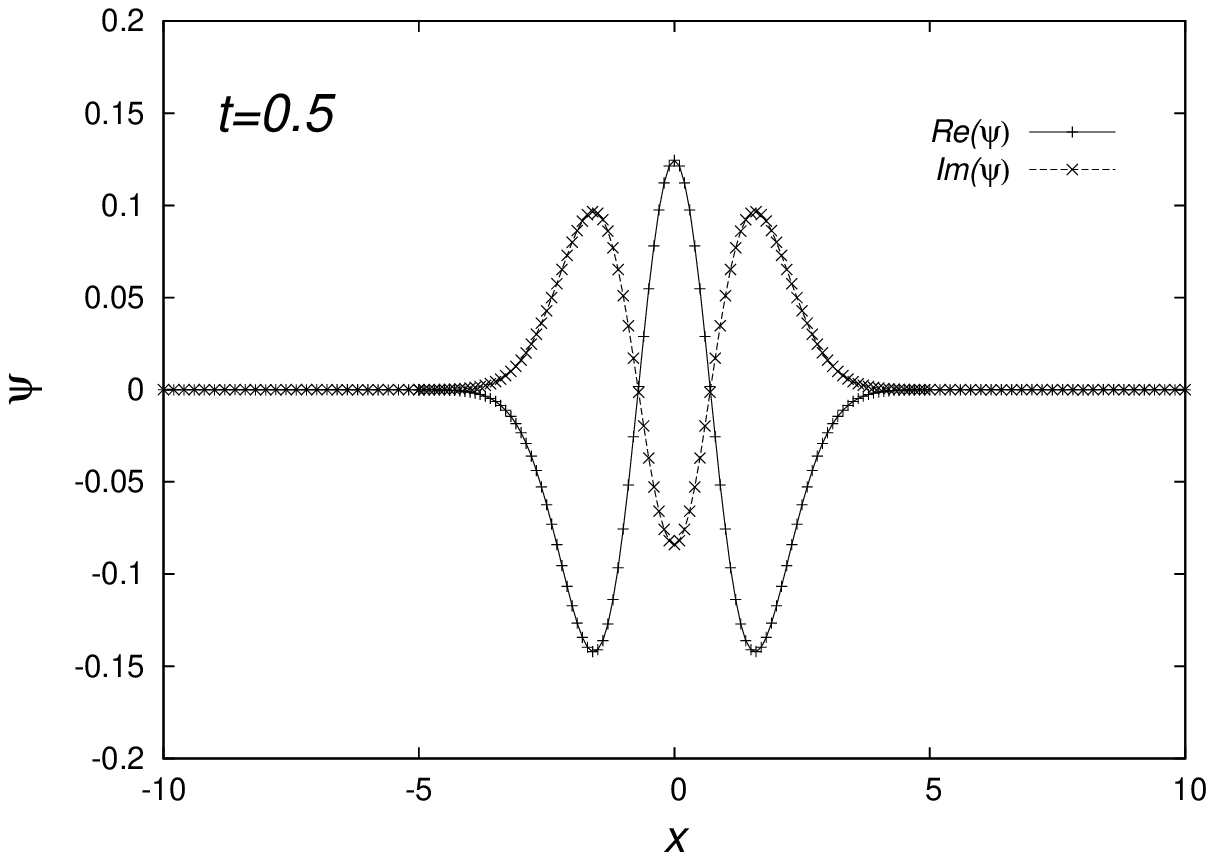}
\includegraphics[width=4.cm]{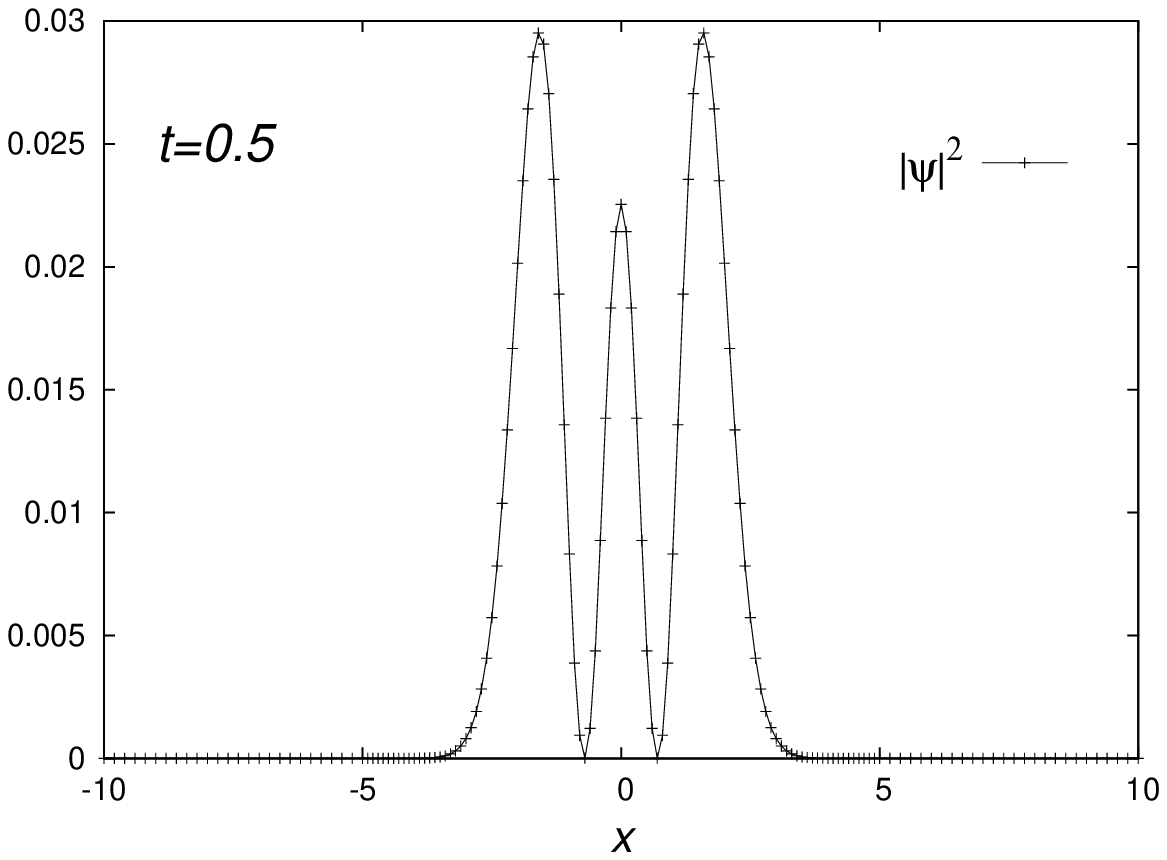}
\caption{\label{fig:HOpsi} We show 1D snapshots of the evolution for the wave function $\Psi_{HO}$ and the density of probability at various times using two refinement levels.}
\end{figure}

\begin{figure}[htp]
\includegraphics[width=7.5cm]{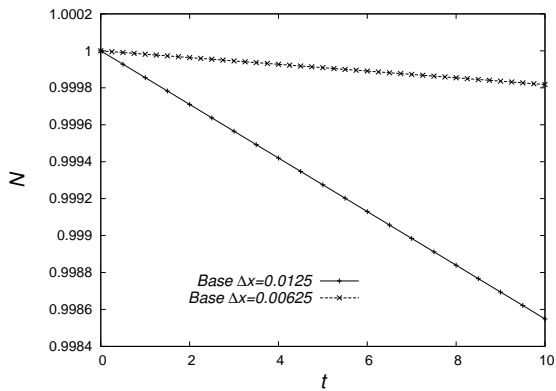}
\caption{\label{fig:HOrhoN} We show $N$ as a function of time for about fifty cycles of the wave function. The dissipation introduced by the evolution algorithm can be reduced when increasing the resolution in time, which in turn implies the conservation of $N$ is more accurate. In our evolutions we choose $\Delta t$ such that it allows long term evolutions and also keep the unitarity constraint accurate.}
\end{figure}

The very first test of an algorithm involving Schr\"odinger equation is the evolution algorithm, which most importantly must be unitary within numerical errors. In order to test this property of the code we choose to evolve the 3D stationary wave function submitted to a harmonic oscillator potential. The code is demanded to evolve the wave function $\Psi$ while the density of probability $|\Psi|^2$ remains time-independent and its integral $N=\int |\Psi|^2 d^3 x$ remains constant in time. The exact normalized  stationary solution is 

\begin{eqnarray}
\Psi_{HO}&=&\sqrt{\frac{1}{\pi 2^{n_x} 2^{n_y} 2^{n_z} n_x ! n_y ! n_z !}}\times \nonumber\\ 
&&e^{-(x^2 + y^2 +z^2)/2} 
H_{n_x}(x)H_{n_y}(y)H_{n_z}(z) \times \nonumber\\
&&e^{-i E_{n_x ,n_y , n_z}t}
\end{eqnarray}

\noindent where the energy is given by $E_{n_x ,n_y , n_z}=n_x + n_y + n_z +1$ and $n_x,~n_y~,n_z$ are the number of nodes along each spatial direction. We then start the evolution of the wave function evaluated at $t=0$ using our algorithms and the FMR driver. In Fig. \ref{fig:HOpsi} we show snapshots of the wave function and the density of probability, showing that even though the wave function evolves, the density of probability remains time-independent. In Fig. \ref{fig:HOrhoN} we show $N$ in time  and  that for even nearly fifty cycles of the wave function $N$ looses about 0.02\%. We show this figure for two different resolutions, showing that the algorithm converges in the continuum limit to a constant value of $N$.

A discussion on the election of evolution algorithm is in turn.
In our implementation we use the MoL instead of the more commonly used implicit Crank-Nicholson algorithm, which in theory is unitary and unconditionally stable for arbitrarily high values of $\Delta t / \Delta x^2$. The stability of the integration with MoL -on the other hand- depends on the stability of the ode solver used; we use here an iterative Cranck-Nicholson integrator, although we also experienced a similar behavior with a third order Runge-Kutta integrator, and in both cases the stability has shown to be safe as long as $\Delta t / \Delta x^2 < 0.5$ in all our runs. Our code works also with the fully implicit Crank-Nicholson method \cite{Choi}, nevertheless we decided to use the MoL because when comparing the accuracy with that of the implicit method, it was comparable for the two algorithms when a similar $\Delta t$ is used, earning little form the appealing stability properties of the implicit method.


\subsection{Test on GPP equilibrium solutions}

\begin{figure}[htp]
\includegraphics[width=7.5cm]{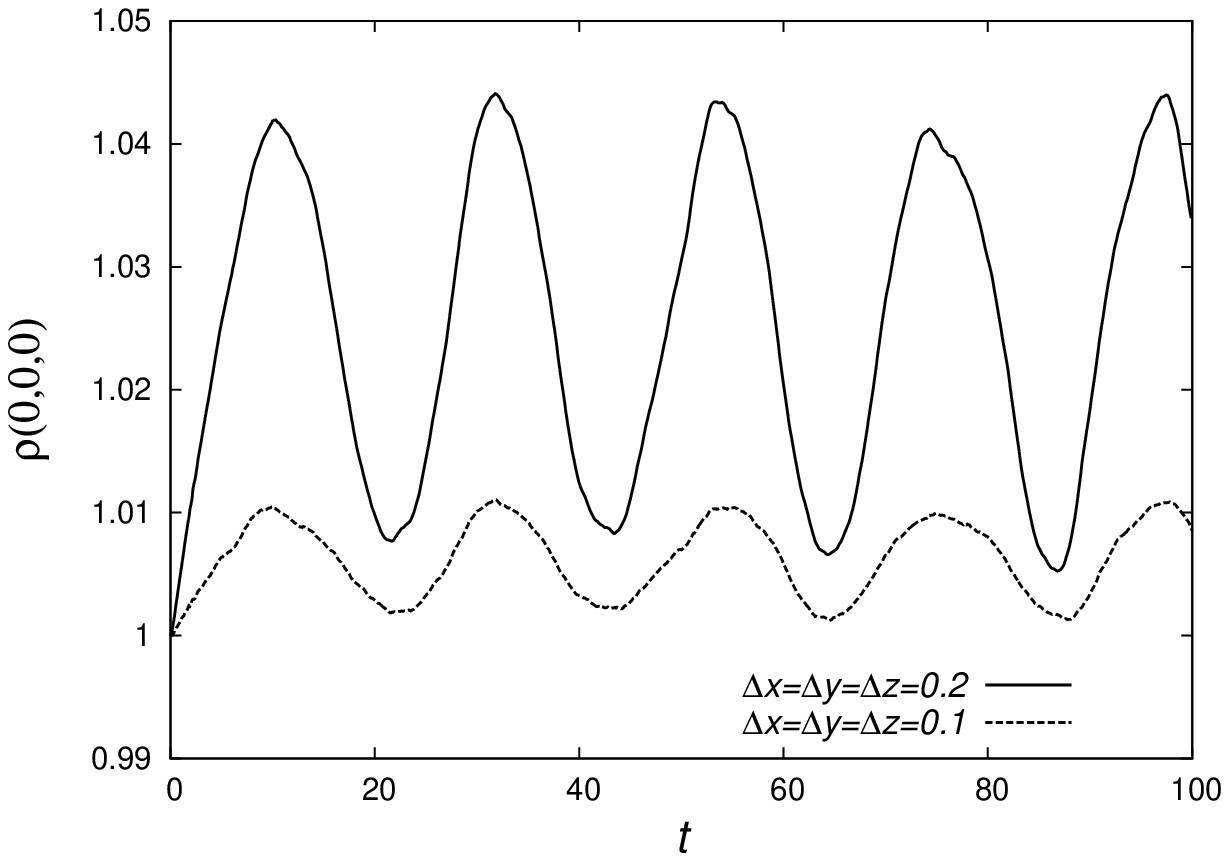}
\includegraphics[width=7.5cm]{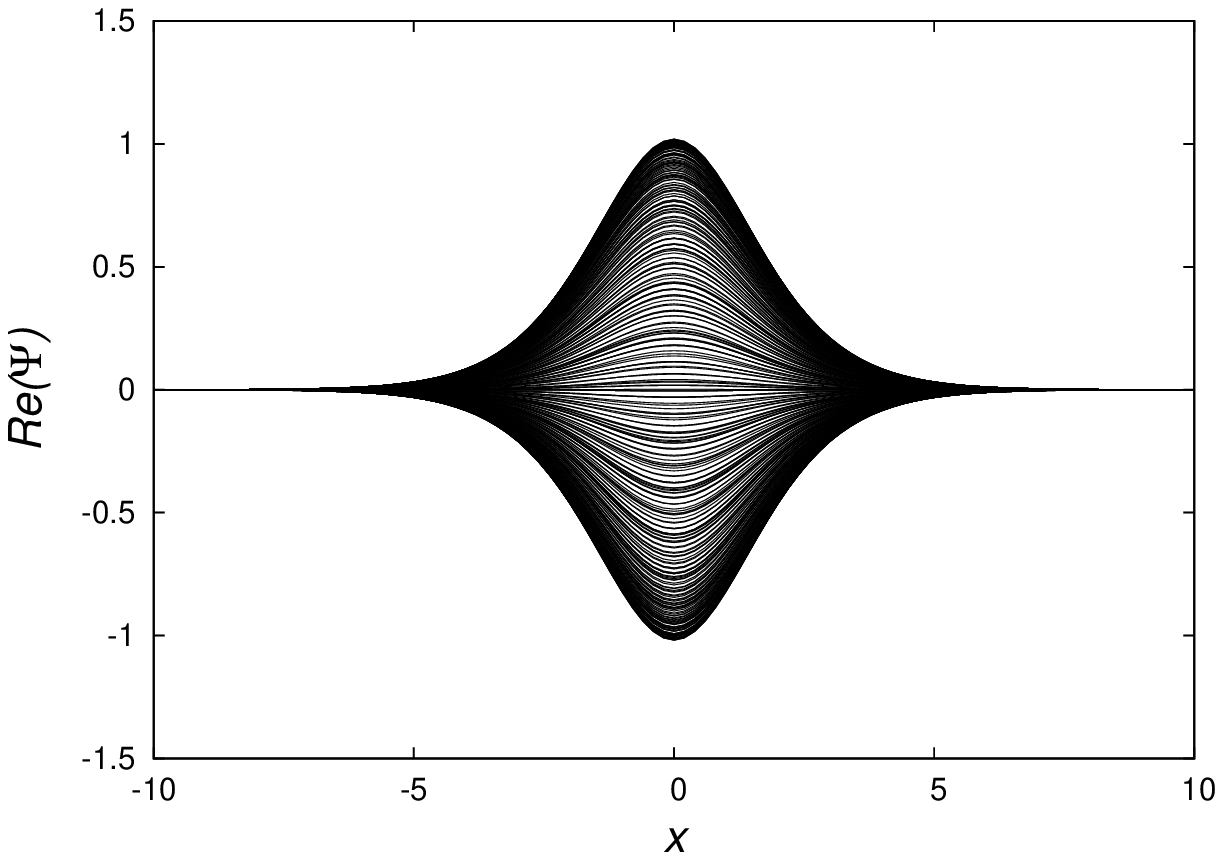}
\caption{\label{fig:test3} (Top) We show the central density of the configuration with $a=0$ (or equivalently, the infinity norm of the density $L_{\infty}(\rho)$) for two different resolutions of our numerical domain $\Delta x =\Delta y = \Delta z = 0.2$ and  $\Delta x =\Delta y = \Delta z = 0.1$. The fact that the departure from $\rho(0,0,0)=1$, in time with the coarse grid is $2^2$ times that with the fine grid indicates the second order convergence of our implementation. (Bottom) we show snapshots of $Re(\Psi)$ and its  considerable dynamics, whereas the density of probability converges to static in the continuum limit.}
\end{figure}

The test of fire of our code consists in the correct evolution of a ground state equilibrium configuration, which incorporates the solution of Poisson equation and therefore includes the accuracy of the elliptic solver implemented. The equilibrium configurations are constructed in spherical coordinates according to \cite{GuzmanUrena2004,GuzmanUrena2006}, which use the normalization that the central value of the density is $\rho(r=0)=1$. We then interpolate this configuration into the 3D domain, and evolve the whole GPP system according to (\ref{eq:SchroNoUnits}-\ref{eq:PNoUnits}). In the continuum limit the density should remain constant in time and space, however discretization errors must converge to zero. What we show in Fig. \ref{fig:test3} is that such errors converge with second order to the result in the continuum limit, that is, the departure of the central value of the density converges to zero with second order. This is consistent with the second order spatial stencils we use to discretize the equations. In order to show that the system has considerable dynamics, we show also in Fig. \ref{fig:test3} snapshots of the wave function in time. 

\begin{figure}[htp]
\includegraphics[width=7.5cm]{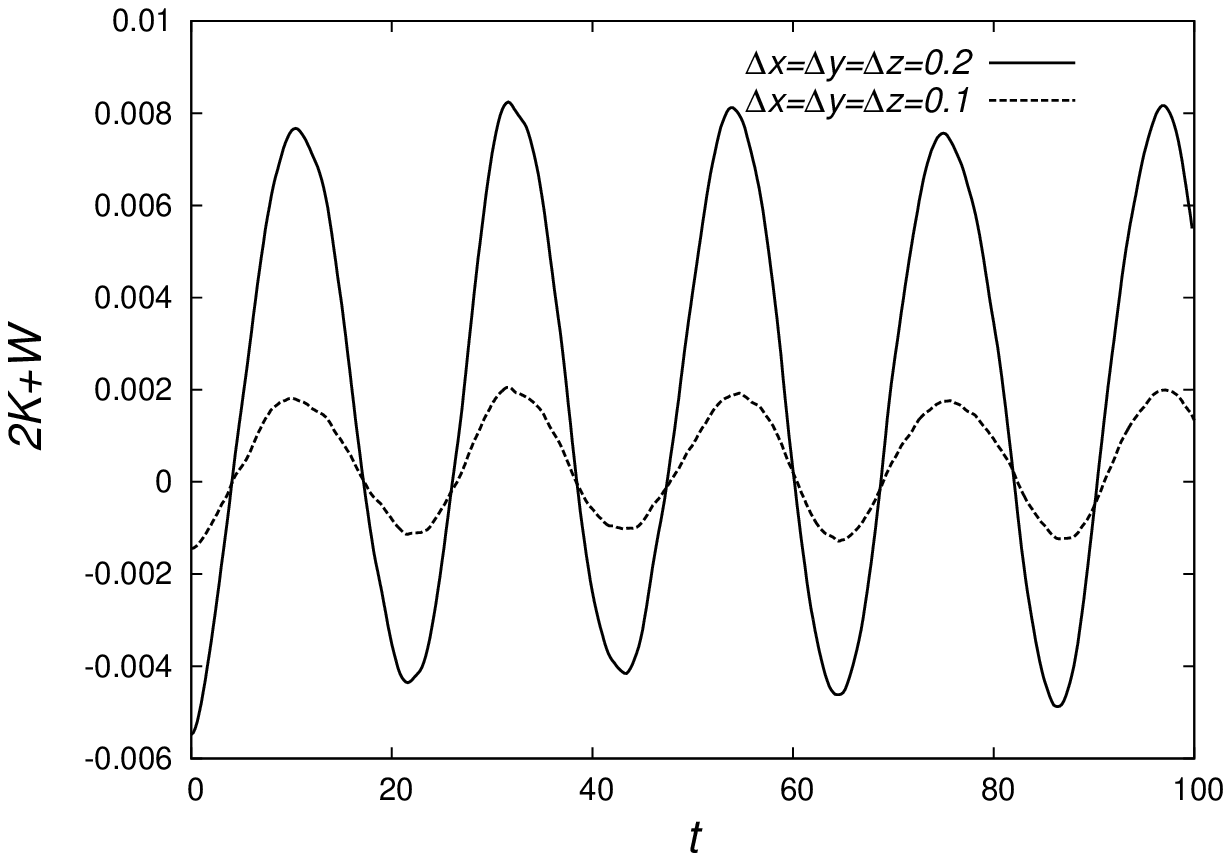}
\includegraphics[width=7.5cm]{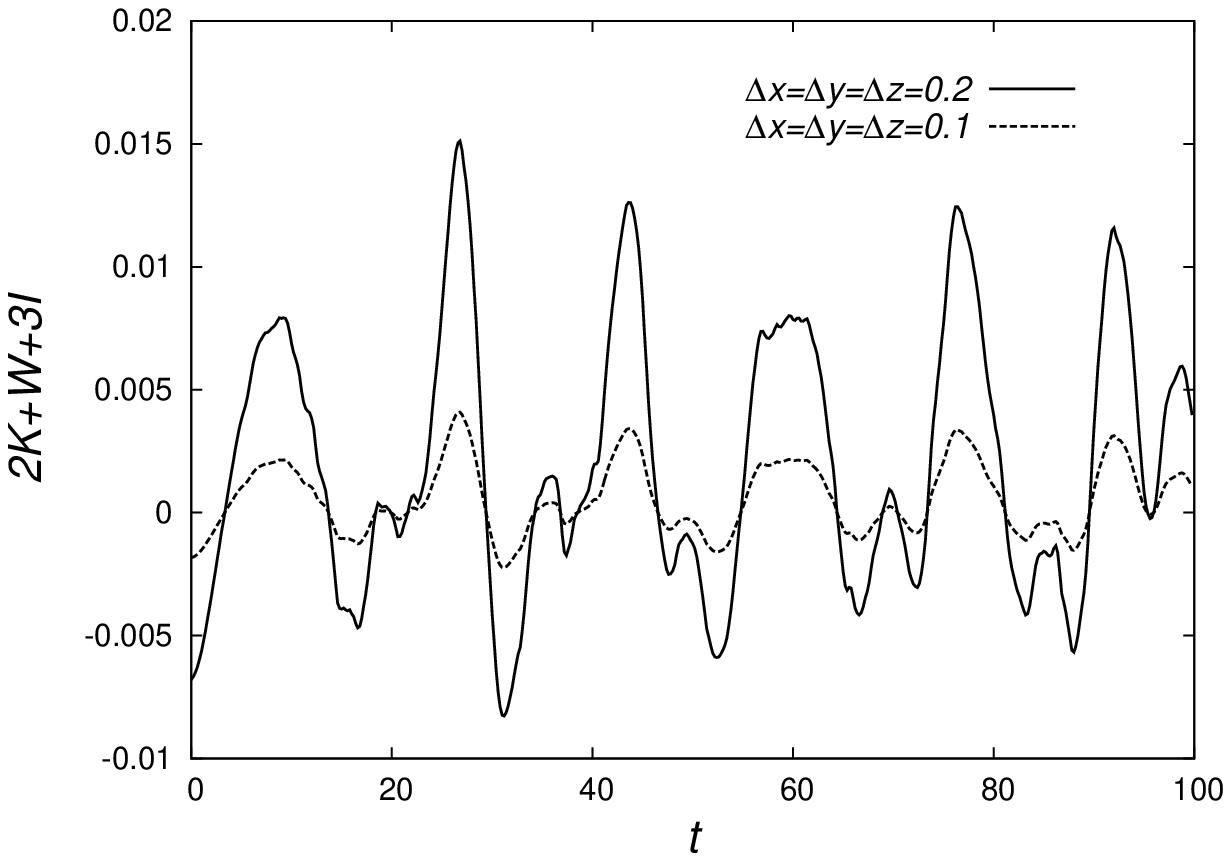}
\caption{\label{fig:test4} We show the $2K+W+3I$ for the standardized spherically symmetric ground state equilibrium solution in spherical symmetry. In the top and bottom panels we show the case $a=0$ and $a=0.5$ respectively. The calculation is made with two resolutions and the second order convergence to zero in both cases indicates the configurations remain virialized in the continuum limit as they should.}
\end{figure}

In Fig. \ref{fig:test4} we show that $2K+W+3I$ converges to zero for the two cases $a=0,~0.5$, which indicates that  the systems remain virialized during the evolution. This test confirms that the elliptic solver used for the  Poisson equation and the evolution algorithm work fine together.

\end{document}